\documentclass[letterpaper,aps,prb,twocolumn,floatfix,amsfonts,amssymb,superscriptaddress]{revtex4-1}
\usepackage{amsmath}
\usepackage{amsthm}
\usepackage{amssymb}
\usepackage{graphicx}
\usepackage{tabularx}
\usepackage{color}
\usepackage{soul}

\newcommand{\comment}[1]{}
\newcommand{\eref}[1]{Eq.~(\ref{#1})}
\newcommand{\Fref}[1]{Figure~\ref{#1}}
\newcommand{\fref}[1]{Fig.~\ref{#1}}
\newcommand{\sref}[1]{Sec.~\ref{#1}}

\newcommand{\D}{\mathrm{d}}
\newcommand{\I}{\mathrm{i}}
\newcommand{\E}{\mathrm{e}}
\newcommand{\acre}[1]{a^{\dagger}_{#1}}
\newcommand{\ades}[1]{a^{\phantom\dagger}_{#1}}
\newcommand{\fcre}[1]{f^{\dagger}_{#1}}
\newcommand{\fdes}[1]{f^{\phantom\dagger}_{#1}}
\newcommand{\fpair}[2]{f^{\dagger}_{#1}f^{\phantom\dagger}_{#2}}

\newcommand{\ccre}[1]{c^{\dagger}_{#1}}
\newcommand{\cdes}[1]{c^{\phantom\dagger}_{#1}}
\newcommand{\hH}{\hat{H}}

\begin{document}
\title{Conductance fingerprint of Majorana fermions in the topological Kondo effect}

\author{Martin R. Galpin}
\affiliation{Department of Chemistry, Physical and Theoretical Chemistry, University of Oxford, South Parks Road, Oxford OX1 3QZ, United Kingdom}
\author{Andrew K. Mitchell}
\affiliation{Department of Chemistry, Physical and Theoretical Chemistry, University of Oxford, South Parks Road, Oxford OX1 3QZ, United Kingdom}
\author{Jesada Temaismithi}
\affiliation{Department of Chemistry, Physical and Theoretical Chemistry, University of Oxford, South Parks Road, Oxford OX1 3QZ, United Kingdom}
\author{David E. Logan}
\affiliation{Department of Chemistry, Physical and Theoretical Chemistry, University of Oxford, South Parks Road, Oxford OX1 3QZ, United Kingdom}
\author{Benjamin B\'eri}
\affiliation{TCM Group, Cavendish Laboratory, University of Cambridge, J. J. Thomson Avenue, Cambridge CB3 0HE, United Kingdom}
\affiliation{School of Physics and Astronomy, University of Birmingham, Birmingham, B15 2TT, United Kingdom}
\author{Nigel R. Cooper}
\affiliation{TCM Group, Cavendish Laboratory, University of Cambridge, J. J. Thomson Avenue, Cambridge CB3 0HE, United Kingdom}

\date{\today}


\begin{abstract}
We consider an interacting nanowire/superconductor heterostructure attached to metallic leads. The device is described by an unusual low-energy model involving spin-$1$ conduction electrons coupled to a nonlocal spin-$1/2$ Kondo impurity built from Majorana fermions.
The topological origin of the resulting Kondo effect is manifest in distinctive non-Fermi-liquid (NFL) behavior, and the existence of Majorana fermions in the device is demonstrated unambiguously by distinctive conductance lineshapes.
We study the physics of the model in detail, using the numerical renormalization group, perturbative scaling and abelian bosonization. In particular, we calculate the full scaling curves for the differential conductance in AC and DC fields, onto which experimental data should collapse. Scattering t-matrices and thermodynamic quantities are also calculated, recovering asymptotes from conformal field theory. We show that the NFL physics is robust to asymmetric Majorana-lead couplings, and here we uncover a duality between strong and weak coupling. The NFL behavior is understood physically in terms of competing Kondo effects. The resulting frustration is relieved by inter-Majorana coupling which generates a second crossover to a regular Fermi liquid. 
\end{abstract}

\maketitle


\section{Introduction}
\label{sec:intro}

Majorana fermions in superconductor heterostructures are presently the most viable candidates for realizing non-Abelian anyons.\cite{read2000paired,*kitaev2001unpaired,FuKane08,*sau2010generic,*alicea2010majorana,*oreg2010helical} These emergent objects are zero-energy modes bound to certain point defects, fractionalising the regular fermionic degrees of freedom. Individual Majoranas can be far apart from each other, so that fermionic modes reconstructed from pairs of Majoranas can be nonlocal in character.\cite{AliMajrev,*BeeMajrev} Qubits encoding the occupation of such modes are topologically protected from local perturbations, and in consequence could find important application within fault-tolerant quantum computation.\cite{nayak2008non}

Several proposals for realizing the Majorana paradigm are the subject of ongoing experimental study, including heterostructures involving semiconductor nanowires with strong spin-orbit coupling,\cite{Mourik25052012,*das2012zero,*Deng2012,rokhinson2012fractional} and those using boundary modes of topological insulators.\cite{Williams2012,Knez2012} Much of this experimental work has focused on demonstrating the existence of Majoranas by detecting the zero-bias anomaly in tunneling conductance, predicted from theory.\cite{LawMaj,*wimmer2011quantum,*Saucond,*pientka2012enhanced} Despite suggestive observations however, unambiguous experimental verification remains elusive because zero-bias peaks can also result from non-Majorana sources.\cite{bagrets2012class,*pikulin2012zero,*liu2012zero,*kells2012nz} 

Compelling evidence for the existence of Majorana fermions in controllable nanodevices should therefore go beyond a direct spectral measurement, probing instead the topological nature of nonlocal qubits. Most theoretical work in this direction relates to phenomena in non-interacting systems.\cite{hassler2010anyonic,*Saunet,*alicea2011non,*vanheck2012coulomb,*vanheckflux2013} However, richer physics could be accessible in systems which can host emergent Majorana particles in the presence of strong interactions between the physical electrons.\cite{Beri2012,Futelep,TsvelikIsing,*crampe2013quantum,Beri2013,*Altland2013,*Zazunov2013}

\begin{figure}[b]
\includegraphics[width=8.5cm]{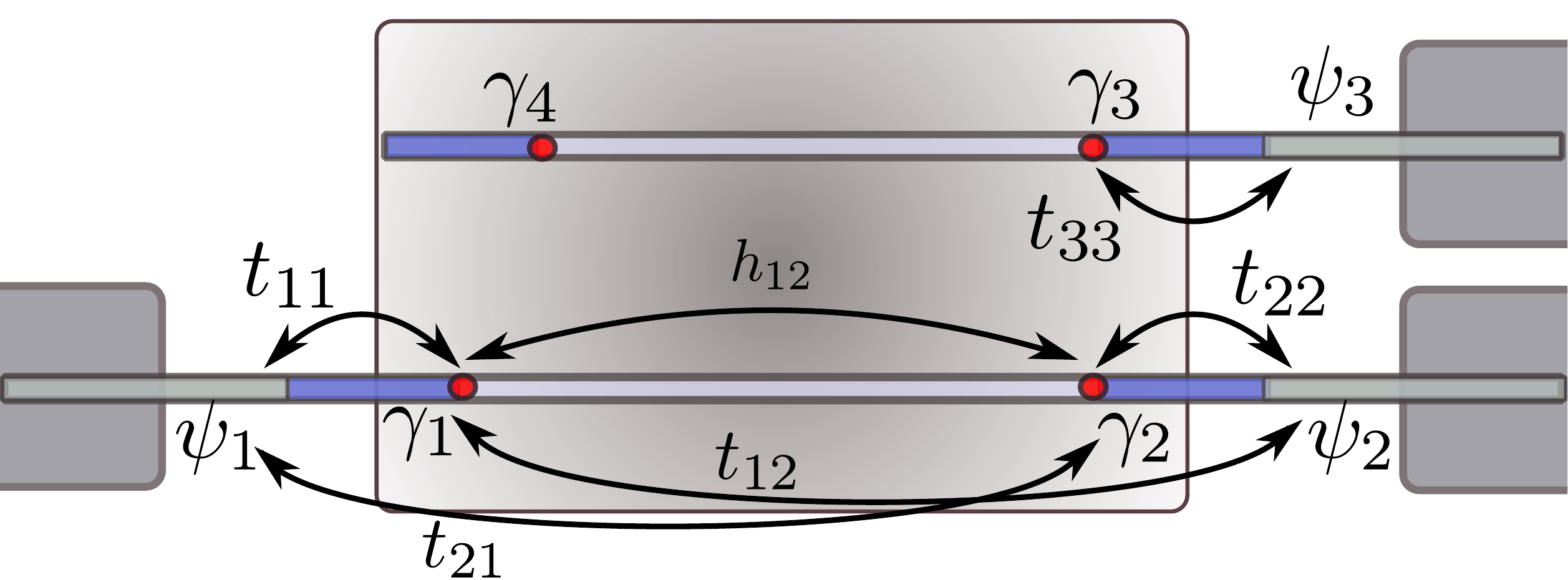}
\caption{\label{fig:MF_schematic} 
The minimal topological Kondo setup and the couplings incorporated in the NRG calculations. The figure shows the sketch of a nanowire realization: the central rectangle is a superconducting island, with two nanowires (horizontal bars). The nanowires turn into gapped superconductors in their central segments, the end of which hosts Majorana fermions (dots). The outer portions of the wires are metallic and are coupled to electron reservoirs (outer rectangles). The wire parts separating metallic and superconducting regions have a depletion gap forming a tunnel barrier.}
\end{figure}

One such scenario was considered recently in Ref.~\onlinecite{Beri2012}: Majoranas at the end of nanowires on an interacting superconducting island produce several topological qubits (see \fref{fig:MF_schematic}). A spin-$1/2$ degree of freedom can be constructed from two such, which may be regarded as a nonlocal `quantum impurity'. Attaching metallic leads to the device allows this state to be probed by conductance measurement and, 
importantly, the spin of the nonlocal impurity is then flipped when an electron is transferred from one lead to the other. This gives rise to an effective exchange coupling between the spin-$1/2$ `impurity' and the conduction electrons (which form a representation of spin-$1$), resulting in a `topological Kondo effect'.\cite{Beri2012}  
The low-energy physics is controlled by renormalization group (RG) flow to an intermediate-coupling non-Fermi-liquid (NFL) fixed point,\cite{Beri2012} itself related\cite{Fabrizio1994,Fabrizio1996} to that of the four-channel Kondo (4CK) model.\cite{Nozieres1980} This leads to distinctive signatures in physical properties, which could be used in experiment to identify clearly the topological Kondo effect, and hence the underlying existence of Majorana fermions in the device.

In this paper we examine this system in detail, going beyond the previous analysis\cite{Beri2012} to calculate the full temperature/energy dependence of physical quantities using the numerical renormalization group (NRG) technique.\cite{Bulla2008} We focus on the differential conductance in AC and DC fields, relevant to experiment. The lineshapes we obtain recover high- and low-energy asymptotes from CFT,\cite{Beri2012} but also contain new information on the entire crossover, which fundamentally encodes the RG flow. 
Experimental data collected on universal temperature/energy scales should collapse to a part of the full scaling curves presented in this paper, allowing experimental verification of the topological Kondo effect.

Our full NRG calculations also confirm a key prediction of Ref.~\onlinecite{Beri2012} that the NFL physics is robust to asymmetric Majorana-lead couplings. This property has important implications for the practical viability of the setup, since fine-tuning is not required. 

A rather complete picture of the complex physics of this system is obtained from analysis of its thermodynamics and scattering t-matrix. Characteristic properties of the NFL fixed point are found at low temperatures, including an unusual $\tfrac{1}{2}\ln(3)$ residual entropy for the Majoranas. Such behavior is similarly obtained at the NFL fixed point of the 4CK model;\cite{Andrei1984,*Tsvelik1985} indeed, asymptotic corrections to fixed point thermodynamics of the form $(T/T_K)^{2/3}$ are common to both models. However, the entire crossover highlights differences, which uniquely fingerprint the topological Kondo effect. This is most clearly seen in the experimentally-relevant case of asymmetric Majorana-lead couplings: here the flow is distinct from that of the 4CK model. Furthermore, we uncover a duality between strong and weak coupling, allowing the Kondo scale at strong coupling to be obtained from simple perturbative scaling performed at weak coupling. In all cases, the NFL fixed point is reached at low energies---unlike the 4CK model, which supports a quantum phase transition to an unscreened local moment state.\cite{Schiller2008}

Finally, we consider relevant symmetry-breaking perturbations which do destabilize the NFL fixed point, generating a crossover to the conventional Fermi liquid state. Going beyond Ref.~\onlinecite{Beri2012}, we identify the coupling \emph{between} different Majoranas to be one such perturbation. While such a coupling is expected to be very small (suppressed exponentially in the inter-Majorana separation), the resulting Fermi liquid crossover scale $T^*_\text{FL}$ must in fact be much smaller than the Kondo scale $T_K$ in order that pristine NFL physics be observed at intermediate temperatures $T^*_\text{FL}\ll T \ll T_K$. In this case, we show that two successive universal crossovers arise: one to the NFL fixed point, and one away from it. Both crossovers are entirely characteristic of the incipient NFL state. 

Given that the Kondo temperature may also be exponentially-small, a real device may have competing $T^*_\text{FL}$ and $T_K$ scales. Here, conductance lineshapes obtained from NRG show more complex behavior, which depends on the ratio $T^*_\text{FL}/T_K$ (a quantity that could be identified for a given experiment). But in all cases, Kondo-enhanced conductance establishes the existence of Majorana fermions in the device.


\section{Derivation of the model}
\label{sec:modelderiv}

We begin by reviewing how the topological Kondo effect arises.\cite{Beri2012} 
The main requirement for the Kondo effect is the coupling of conduction electrons to an impurity with a degenerate ground state. In the topological Kondo context, the impurity is constructed using a superconducting island with Majorana fermions (see \fref{fig:MF_schematic}).   The island is of mesoscopic size, characterised by a charging term $\hH_c=E_c\left(\hat N-\frac{q}{e}\right)^2$, where $\hat N$ is the number operator for the island electrons and $E_c$ is the charging energy. The ground state will be degenerate if there are at least four Majoranas. Focusing on this minimal case (see also Fig.~\ref{fig:MF_schematic}), if the Majorana wavefunctions do not overlap, each state with a given $N$, and thus the ground state in particular,  is twofold degenerate: the four Majoranas combine into two zero energy fermions with a fixed overall parity (their occupation can be changed only by transferring Cooper pairs to/from the superconducting condensate).\cite{AliMajrev} It is this degeneracy that leads to an effective spin degeneracy for our Kondo model. If one includes the overlap of  Majorana fermions $\gamma_j$ and $\gamma_k$, the degeneracy is not exact: a ``Zeeman" splitting that is exponentially small in the Majoranas' separation arises, given by the term $\hat H_M$ considered below.

The topological Kondo system is obtained by coupling the island to conduction electrons. In what follows, we focus on the minimal setup sketched in Fig.~\ref{fig:MF_schematic}: we couple three of the four Majorana fermions to single-channel leads of effectively spinless conduction electrons (generated in practice by the application of a large magnetic field or spin-orbit coupling\cite{Beri2012}). The conduction electron spin densities, vital for any Kondo effect,  will arise from nonlocal combinations of electron operators of different leads.\cite{Beri2012} 

Working at energy scales much below the superconducting gap,  and also much below the energies of any non-Majorana sub gap excitations, the physics 
is described by the Hamiltonian
\begin{equation}
\label{eq:Horig}
\hH=\hat{H}_\text{leads}+\hH_\text{isl}+\hH_\text{tun}
\end{equation}
where 
 \begin{equation}
 \hH_\text{isl}=\hH_c+\hH_M,
 \end{equation}
\begin{equation} 
\label{eq:leads_physical}
\hat{H}_{\text{leads}} = \sum_{k, j}\epsilon^{\phantom{\dagger}}_{k} \acre{k j}\ades{k j} \; , 
\end{equation}
\begin{equation}
\hH_M=h_{jk}\ i\gamma_j \gamma_k\;,
\end{equation}
and
\begin{equation} 
\label{eq:Htun}
\hat{H}_\text{tun}=\exp(\I \hat\phi/2) \sum_{i,j} t^{\phantom{\dagger}}_{ij} \gamma^{\phantom{\dagger}}_{i}\psi_{j} + \text{H.c.}  
\end{equation}
Here $\hat{H}_{\text{leads}}$ is the Hamiltonian of conduction electrons, with $a^\dagger_{k,j}$ creating scattering states (standing waves)  of ingoing momentum $k$ in lead $j$. The term $\hat{H}_\text{tun}$ describes the low energy coupling between the leads and the island. The operators $\psi_{j}$ correspond to localized conduction electron orbitals at the end of each physical wire,
\begin{equation}
\label{eq:cb_locorb}
\psi_{j} = N_\mathrm{orb}^{-1/2}\sum_{k} \ades{k j} \; ,
\end{equation}
and the phase exponential $\exp(\pm \I \hat\phi/2)$ is also an operator, ensuring charge conservation by changing $N\rightarrow N\pm 1$. 
To obtain $\hat{H}_\text{tun}$, one expresses the electron operator on the island as $\psi_S(x)=\exp(-\I\hat\phi/2)[\sum_j\xi_j(x)\gamma_j+\psi_{>,S}(x)]$ [where $\xi_j(x)$ are the Majorana wavefunctions and $\psi_{>,S}(x)$ is the piece of the electron operator with positive energy BCS excitations], and writes down the usual hopping terms $\sim \psi_S^\dagger(x_j)\psi_j+\text{h.c.}$ (where $x_j$ is the tunneling point in terms of the coordinates on the island). Below the energy scale of the superconducting gap, as discussed above, the $\psi_{>,S}(x)$ piece can be neglected, which leads to Eq.~\ref{eq:Htun}.\cite{Futelep} 

The key difference between the topological Kondo problem of Ref.~\onlinecite{Beri2012} and Eq.~\eqref{eq:Horig} is that the former includes only the local couplings $t_{ii}$, while here we also consider the nonlocal $t_{i\neq j}$ and $h_{jk}$ terms. Of course, due to the well-localized nature of the Majorana wavefunctions $\xi_j(x)$, these are exponentially suppressed with the inter-Majorana distance, $|t_{i\neq j}|, |h_{jk}|\ll t_{ii}$. (We chose the phase of the electron operators so that $t_{ii}>0$.) At the lowest energy scales, as we will show, they will nevertheless lead to interesting, qualitatively new features. 

On energy scales much lower than the charging energy $E_c$, if $|t_{ij}|\ll E_c$ one can simplify the model by a standard Schrieffer-Wolff transformation.\cite{Hewson1997} Here it is convenient to assume that the charging term of $\hat H_c$ is tuned to the middle of a Coulomb blockade valley\footnote{Moving away from the middle of the Coulomb blockade valley would generate non-universal (but RG-marginal) potential scattering terms in $\hat H_\text{SW}$.\cite{Beri2012} We have checked numerically that that the non-Fermi-liquid fixed point discussed in \sref{sec:symm} is robust to their inclusion; their effect on physical properties such as the finite-$T$/finite-$\omega$ differential conductance is left for future work.} where $\langle\hat H_c\rangle = 0$ for some $N$; by considering virtual excitations to states of island charge $N\pm 1$, one obtains
\begin{equation}
\label{eq:Heffwt}
\hat H_{\text{SW}}=H_{\text{leads}}+i h_{jk}\gamma_{j}\gamma_{k}+\sum_{k\neq q}\frac{2t_{qp}^{*}t_{kl}}{E_{c}}\gamma_{k}\gamma_{q}\psi_{p}^{\dagger}\psi_{l}
\end{equation}
in which repeated indices are summed over. (We add that $h_{jk}$ type contributions are also generated by the $t_{i\neq j}$ couplings; these have been absorbed into the second term.) 

Using now that the Majorana bilinears $\gamma_1\gamma_2$, $\gamma_2\gamma_3$ and $\gamma_3\gamma_1$ (the only independent ones due to the zero-mode parity constraint) form a  spin-$\tfrac{1}{2}$ operator $\hat S_{\alpha}$ via
\begin{equation}
\label{eq:imp_mf}
\boldsymbol{\hat{S}}=-\frac{\I}{4}\boldsymbol{{\gamma}}\times \boldsymbol{{\gamma}} \; ,
\end{equation}
the effective model can be written in the form,  $\hat{H}_{\text{SW}}=\hat{H}_{\text{leads}}+\hat{H}_{K}+\delta\hat{H}$. The conduction electron Hamiltonian $\hat{H}_{\text{leads}}$ is given in \eref{eq:leads_physical}, and the Kondo coupling term is\cite{Beri2012} 
\begin{equation} 
\label{eq:model}
\hat{H}_{K} = \sum_{\alpha} \lambda_{\alpha\alpha} \hat{S}_{\alpha} \hat{I}_{\alpha} \; ,
\end{equation}
where $\hat{I}_{\alpha}$ is a spin-$1$ operator for the lead electrons,
\begin{equation}
\label{eq:cb_spin1}
\hat{I}_{\alpha} = \I \sum_{a,b} \epsilon^{\phantom{\dagger}}_{\alpha b a} \psi^\dagger_{a}\psi_{b} \; ,
\end{equation}
and the positive couplings $\lambda_{\alpha \alpha}\sim \epsilon_{\alpha j k} t_{jj} t_{kk}/E_c$. For $\delta \hat{H}=0$, \eref{eq:model} can thus be viewed as an (anisotropic)  Kondo model involving a spin-$\tfrac{1}{2}$ `impurity' coupled to spin-$1$ conduction electrons.\cite{Fabrizio1994,Beri2012}

Non-local couplings $t_{i\neq j}$, $h_{jk}$, to leading order in exponentially small quantities, generate the term
\begin{equation} 
\label{eq:pert}
\delta\hat{H} = \sum_{\alpha\neq\beta}\lambda_{\alpha\beta}\hat{S}_{\alpha}\hat{I}_{\beta}+\sum_{\alpha,\beta}\lambda_{\alpha\beta}^{\prime}\hat{S}_{\alpha}\hat{I}_{\beta}^{(2)}+h_{\alpha}\hat S_\alpha \;.  
\end{equation}
Here $\hat{I}_{\beta}^{(2)}=\sum_{ab}\psi^\dagger_a [J^{(2)}_\beta]_{ab}\psi_b$ are the five components of a spin-2 density where $J^{(2)}_\beta$ are elementary symmetric real matrices, and the real coupling constants are $\lambda_{\alpha \beta}\sim \lambda^{\prime}_{\alpha\beta} \sim |t_{i\neq j}|t_{jj}/E_c$, and $h_\alpha\sim \epsilon_{\alpha ij}h_{ij}$. 


\subsection{Simplification to the axial-symmetric limit}
\label{sec:axsymm}

While the full model described by \eref{eq:model}--\eref{eq:pert} can in principle be treated by the NRG (to be described later), the calculations are computationally rather expensive. There are two reasons for this: only the total charge is a conserved quantum number, and the Hamiltonian matrix has complex elements. The model as written also contains a large number of parameters, and one cannot hope to examine its full parameter space exhaustively.

We therefore adopt the following simplifications. For much of the paper, we focus on $\delta \hat H = 0$, to identify and understand the universal physics arising from the topological Kondo effect of \eref{eq:model}. To simplify the calculations, we now assume $t_{11}=t_{22}$, and imaginary $t_{12}=-t_{21}$. This leads to a residual `axial symmetry' around axis $3$, as will be discussed in due course, which allows the calculations to be performed with real matrix elements and exploiting the conservation of an additional, overall $S_z$ quantum number.

To obtain a handle on the key effects of the perturbation $\delta \hat H$, one can focus on the non-local couplings arising from the exponentially small overlap between Majoranas $\gamma_1$ and $\gamma_2$ alone. This simplifies the model considerably, as it can then be shown that the first term in Eq.~\eqref{eq:pert} is then absent. We leave a study of the more general case to future work; our expectation is that the remaining perturbations we keep in $\delta \hat H$ are sufficient to understand the essential effects of non-local couplings between the Majoranas: namely that if the non-local couplings are made sufficiently small, the universal non-Fermi-liquid physics of the model persists above a low-energy crossover scale set by the size of $\delta \hat H$ (see \sref{sec:breakab}).

With these simplifications in place, we employ a unitary transformation of the lead operators to a basis labeled by the conduction electron spin projection, $m=-1, 0, +1$, viz, 
\begin{subequations}
\label{eq:lead_trasnform}
\begin{align}
\cdes{k,-1} &= \frac{1}{\sqrt{2}}\left(\ades{k 1} + \I\ades{k 2}\right) \; ,\\
\cdes{k,0} &= \ades{k3}\label{eq:3trans} \; ,\\
\cdes{k,+1} &= \frac{1}{\sqrt{2}}\left(-\ades{k 1} + \I\ades{k 2}\right) \; .
\end{align}
\end{subequations}
The lead Hamiltonian then follows simply as $\hat{H}_{\text{leads}}=\sum_{k, m}\epsilon^{\phantom{\dagger}}_{k} \ccre{k,m}\cdes{k,m}$. Localized orbitals in the new basis are defined as
\begin{equation}
\label{eq:f-orb}
f_m = N_\mathrm{orb}^{-1/2}\sum_k \cdes{k, m} \;, 
\end{equation}
in terms of which the spin-$1$ ladder operators $\hat I^\pm = \hat I_x \pm \I \hat I_y$ of the lead electrons are 
\begin{subequations}
\label{eq:spin1}
\begin{align}
\hat{I}^+ &= \sqrt{2}(\fcre{0}\fdes{-1} + \fcre{+1}\fdes{0}),\\
\hat{I}^- &= \sqrt{2}(\fcre{0}\fdes{+1} + \fcre{-1}\fdes{0}),\\
\hat{I}_z &= \fcre{+1}\fdes{+1}-\fcre{-1}\fdes{-1}.
\end{align}
\end{subequations}
It is then straightforward to show that under the above transformation, the Kondo model including the perturbations takes the form
\footnote{If one takes $t_{i\neq j}=0$, the coupling constants in \eref{eq:heff} are simply related to those of \eref{eq:model} by $J_{\perp}^a=J_{\perp}^b=\lambda_{11}=\lambda_{22}$, and $J_z=\lambda_{33}$. Non-local couplings $t_{i\neq j}\ne 0$ generate the $h_z$ term and unequal $J_\perp^a \ne J_\perp^b$, and then the relationship between the two sets of coupling constants becomes more complicated.} 
\begin{equation}
\begin{split}
\label{eq:heff}
\hat{H}_{\text{eff}} = \hat{H}_{\text{leads}} &+ \frac{J_\perp^a}{\sqrt{2}}\left[S^+\fcre{0}\fdes{+1} + S^-\fcre{+1}\fdes{0}\right]\\
&+ \frac{J_\perp^b}{\sqrt{2}}\left[S^+\fcre{-1}\fdes{0} + S^-\fcre{0}\fdes{-1}\right] \\
&+ J_z S_z(\fcre{+1}\fdes{+1}-\fcre{-1}\fdes{-1})+h_{z}\hat S_z,
\end{split}
\end{equation}
with $\hat S^\pm = \hat{S}_x \pm \I \hat{S}_y$ and $\hat{S}_z$ spin-$\tfrac{1}{2}$ operators for the `impurity' as before. 

The dependence of the four coupling constants in \eref{eq:heff} on the original model parameters is generally non-trivial. Since the aim of this work is to examine the universal physics of the model and understand the basic effect of non-local couplings between Majoranas, we shall treat the coupling constants of \eref{eq:heff} as the bare model parameters of interest. With this in mind, we make one final simplification: we set $h_z = 0$. Since a Zeeman term is associated with the same effective time-reversal symmetry breaking as setting $J_\perp^a\ne J_\perp^b$, the basic effect of this symmetry breaking can be probed by considering the latter alone.   

To calculate thermodynamic and dynamical properties of \eref{eq:heff}, we employ the NRG, a powerful non-perturbative method which yields numerically-exact results over a wide range of temperature/energy scales. The general procedure for calculating thermodynamics is explained fully in the review of Ref.~\onlinecite{Bulla2008}, to which we refer the reader for further details. 

The key approximation in NRG is a logarithmic discretization of the conduction electron densities of states. Since we focus on the universal physics of the model here, it suffices to consider equivalent symmetric bands, each with constant density of states $\rho = 1/(2D)$ over a bandwidth $2D$. These are discretized and transformed into semi-infinite 1d tight-binding `Wilson chains', 
\begin{equation}
\label{eq:discleads}
\hat{H}_{\text{leads}}^{\textit{disc}} = \sum_m \sum_{p=0}^{\infty} \left ( t^{\phantom{\dagger}}_{p} \fcre{m,p}\fdes{m+1,p} + \text{H.c} \right ) \;,
\end{equation}
where the impurity couples only to the `zero-orbital' $\fdes{m,0} \equiv \fdes{m}$, as defined in \eref{eq:f-orb}. The logarithmic discretization means that the Wilson chain hoppings $t_p$ decrease exponentially down the chain, rendering the problem amenable to an iterative solution in which high-energy states are successively discarded as more Wilson chain orbitals are added.\cite{Bulla2008} 
Dynamics are calculated within the complete Anders-Schiller basis,\cite{Anders2005} using the `full density matrix' approach.\cite{Weichselbaum2007,*Peters2006}  In practice we exploit the $U(1)$ symmetries of \eref{eq:heff} (overall charge and $S_z$ conservation). We use a discretization parameter $\Lambda = 3$, and retain at most 10000 states at each iteration. The results of 8 separate calculations with different discretization `slide parameter' are combined to obtain highly accurate dynamics.\cite{Oliveira1994}


\section{Fixed points and symmetries}
\label{sec:symm}
Before presenting numerical results, we first identify and discuss the fixed points of the model, \eref{eq:heff}, and consider the RG flows between them. We begin with the axial symmetric limit $J_\perp^a = J_\perp^b$, where an intermediate coupling, non-Fermi-liquid fixed point is stable. Further insight into this fixed point can then be gained by considering the behavior of the model when $J_\perp^a \ne J_\perp^b$: the intermediate coupling fixed point arises from a competition between two Kondo effects, as explained below. 

At high energies, the physics of the model is controlled by the local moment (LM) fixed point, obtained by setting $J_{\perp}^a=J_{\perp}^b=J_z=0$ in \eref{eq:heff}. At the fixed point itself, the spin $\mathbf{\hat S}$ (the `impurity') decouples from the three conduction electron channels, to give\cite{Hewson1997} an `impurity entropy' $S_{\text{imp}}=\ln 2$ and Curie law magnetic susceptibility $T\chi_{\text{imp}}=\tfrac{1}{4}$ (we use units where $k_B\equiv 1$ and $(g\mu_B)^2\equiv 1$ throughout). Near the fixed point, antiferromagnetic exchange coupling is (marginally) relevant, and its effect can be understood using Anderson's poor man's scaling\cite{Anderson1970,Beri2012} (i.e. perturbative RG) to obtain flow equations for the renormalized couplings as the effective bandwidth/energy scale is reduced. Defining dimensionless running couplings $j_{\perp}=\rho J_{\perp}^a=\rho J_{\perp}^b$, $j_{z}=\rho J_z$, and with $\tilde{D}$ the running UV cutoff, second-order poor man's scaling gives
\begin{equation}
\label{eq:pms}
\frac{d j_{\perp}}{d \ln \tilde{D}} = - j_{\perp} j_{z} \qquad ; \qquad \frac{d j_{z}}{d \ln \tilde{D}} = - j^2_{\perp} \;.
\end{equation}
These equations are precisely those obtained for the spin-$\tfrac{1}{2}$ anisotropic Kondo model, with a single spinful conduction electron channel.\cite{Hewson1997}
The initial RG flow is therefore similar to that of the regular Kondo problem: weak antiferromagnetic bare couplings begin to grow as the temperature/energy scale is reduced, showing up in physical quantities such as the conduction electron t matrix and conductance as slow inverse-logarithmic tails at high energies.\cite{Hewson1997,Dickens2001}

In the antiferromagnetic one-channel Kondo problem, it is well known\cite{Hewson1997} that RG flows tend ultimately to a stable, isotropic strong coupling (SC) fixed point, describing the Kondo singlet at very low energies. [This conclusion is naturally beyond the scope of the perturbative scaling analysis, but has been established by exact methods including NRG\cite{Bulla2008} and the Bethe ansatz.\cite{tsvelick1983exact,Andrei1983}] 
An isotropic SC fixed point also exists for \eref{eq:heff}, obtained by setting $J_\perp^a = J_\perp^b = J_z = \infty$ in \eref{eq:heff} and $t_0=0$ in \eref{eq:discleads} (which decouples the three leads subject to a $\pi/2$ phase shift). But in contrast to the conventional Kondo model, this SC fixed point is unstable. The reason is that the decoupled subsystem comprising the impurity and the three $\fdes{m}$ `zero orbitals' has degenerate ground states with local charge $n_\mathrm{SC} = \sum_m \langle\fcre{m}\fdes{m}\rangle = 1$ and $2$, which are connected by a relevant perturbation of the form
\footnote{We note\cite{Krishnamurthy1980} that the same perturbation connects degenerate charge states of the `free orbital' fixed point in the regular Anderson impurity model}
\begin{equation}
\delta \hat H_{\text{SC}} = \sum_m \fcre{m,0} \fdes{m,1} + \mathrm{H.c.} \;.
\end{equation}
In the conventional Kondo model, there is no such internal structure at the SC fixed point: the strong coupling state is an `inert' singlet and no such relevant perturbations around the fixed point are possible. 

With both LM and SC fixed points unstable when $J_\perp^a = J_\perp^b$, one anticipates the existence of a stable fixed point at intermediate coupling. 
This intermediate coupling fixed point was shown to be stable within a form of the Toulouse limit for a model with axial symmetry.\cite{Fabrizio1994} A full CFT\cite{affleck1990current,*affleck1991kondo,*affleck1991critical,*Affleck1993} analysis of the problem was recently performed,\cite{Beri2012} establishing that this overscreened NFL fixed point is robust to breaking exchange isotropy ($\lambda_{11}\neq\lambda_{22}\neq\lambda_{33}$) or, equivalently, to asymmetries of the local couplings $t_{jj}$ in the original model of \eref{eq:Horig}.

Further insight into the model can be obtained by noting that the spin sector of the model, \eref{eq:heff}, is identical to that of the four-channel Kondo (4CK) model.\cite{Fabrizio1994,Fabrizio1996,Sengupta1996} The physics near the intermediate coupling fixed point therefore has many common features with that of the 4CK effect. The latter has been studied using the Bethe ansatz\cite{Andrei1984,*Tsvelik1985} and boundary CFT,\cite{affleck1990current} which predict for example a residual impurity entropy $S_{\text{imp}}=\ln \sqrt{3}$ and NFL low-temperature behavior such as a divergent susceptibility $\chi_{\text{imp}}\sim T^{-1/3}$. As these thermodynamic quantities originate from the spin sector, the same behavior is expected for the topological Kondo effect studied here. But crucially, the remaining sectors of the two models are distinct. The realization of the NFL fixed point with a spin-$1$ conduction band changes certain key physical properties, as we will illustrate through our calculation of the scattering $t$-matrix (see \sref{sec:tmatrix}). Moreover, although the fixed points of \eref{eq:heff} and the 4CK model are similar, the RG flows \emph{between} them are also quite different, as we shall discuss in \sref{sec:axial}.\\

Upon including non-local couplings $t_{i\neq j}$, we perturb the fixed point studied in Ref.~\onlinecite{Beri2012} in an as-yet unexplored manner. These perturbations lead to qualitative changes: since they introduce complex couplings in Eq.~\eqref{eq:Heffwt}, they break the effective time-reversal invariance\cite{Beri2012} responsible for the stability of the NFL fixed point. But if these couplings are sufficiently small (see \sref{sec:breakab} for a precise definition of `small'), the NFL behavior will persist down to an energy scale $T_\mathrm{FL}^*$, before the symmetry-breaking becomes important and deviations occur. 

In terms of the simplified model under consideration in this work, \eref{eq:heff}, the non-local couplings manifest themselves in two ways. As discussed in \sref{sec:modelderiv}, they introduce an effective magnetic field $h_z$ (neglected here, as explained in \sref{sec:axsymm}) and they lead to $J_\perp^a \ne J_\perp^b$. To understand the effect of the latter further, we can rewrite \eref{eq:heff} as 
$\hat{H}_{\text{eff}} = \hat{H}_{\text{leads}} +\hat{H}_K^a +\hat{H}_K^b$, with 
\begin{eqnarray}
\label{eq:hintrewrite}
\hat{H}_K^a &=\frac{J_\perp^a}{\sqrt{2}}\left[S^+\fcre{0}\fdes{+1} + S^-\fcre{+1}\fdes{0}\right] + J_z^a S_z(\fcre{+1}\fdes{+1}-\fcre{0}\fdes{0}), \;\;\;\;\;\;\;\; \\
\hat{H}_K^b &=  \frac{J_\perp^b}{\sqrt{2}}\left[S^+\fcre{-1}\fdes{0}
+ S^-\fcre{0}\fdes{-1}\right] + J_z^b S_z(\fcre{0}\fdes{0}-\fcre{-1}\fdes{-1}), \;\;\;\;\;\;\;\;
\end{eqnarray}
and $J_z^a = J_z^b = J_z$.  In this form the model has a simple physical interpretation. The terms $\hat{H}_K^a$ and $\hat{H}_K^b$ each describe a spin-$\tfrac{1}{2}$ anisotropic Kondo model, with spin-$\tfrac{1}{2}$ conduction electrons. In $\hat{H}_K^a$, channels $m=0$ and $+1$ can be thought of as the `$\uparrow$' and `$\downarrow$' spins of a conventional Kondo model; while in $\hat{H}_K^b$ these roles are played by channels $m=0$ and $-1$. Crucially, channel $m=0$ plays a different role in $\hat{H}_K^a$ than it does in $\hat{H}_K^b$ --- it is therefore not possible to combine channels $m=\pm 1$ into a single `$\uparrow$' channel and have channel $m=0$ acting as the `$\downarrow$' spin in both terms (this is merely a statement that Kondo models with spin-$\tfrac{1}{2}$ and spin-$1$ conduction electrons are inequivalent).

When $J_{\perp}^a > J_{\perp}^b$, then $J_{\perp}^a$ grows under renormalization faster than $J_{\perp}^b$, and the system is driven to an isotropic strong coupling fixed point. The impurity spin-$\tfrac{1}{2}$ is exactly Kondo-screened by the $m=0$ and $+1$ channels of $\hat{H}_K^a$, with $J_\perp^a/\sqrt{2} = J_z^a \to \infty$. Meanwhile, the third channel, $m=-1$ decouples, $J_\perp^b = J_z^b \to 0$. By contrast, when $J_{\perp}^b > J_{\perp}^a$, the Kondo effect takes place with the $m=0$ and $-1$ channels, with $m=+1$ now decoupling.

The critical physics of the model near $J_\perp^a = J_\perp^b$ can thus be understood physically in terms of competing Kondo effects in different channels, depending on whether $J_{\perp}^a \lessgtr J_{\perp}^b$. On tuning precisely to $J_{\perp}^a = J_{\perp}^b$ (i.e.~$\delta \hat H =0$), one realizes the stable NFL fixed point of the model: by symmetry, neither $m=-1$ nor $m=+1$ can individually decouple, with the result that the impurity is `overscreened'. This competition between strong coupling Kondo-screened states, and the accompanying emergence of NFL physics, similarly arises in the multichannel Kondo\cite{Nozieres1980,Mitchell2013} and multi-impurity Kondo\cite{Jones1988,*Jones1991,Affleck1992,*Affleck1995,Mitchell2012,Mitchell2013} problems.

The quantum phase transition arising on tuning $J_{\perp}^a / J_{\perp}^b$ is explored further in \sref{sec:breakab}.


\section{NRG Results for the channel-isotropic limit}
\label{sec:resiso}
We begin by examining the physics when $J^a_\perp = J_{\perp}^b=J_z \equiv J$ [i.e. equal local couplings, $\lambda_{\alpha\alpha} = \lambda$ in \eref{eq:model}]. Here the model has full $SU(2)$ spin symmetry, with each screening channel coupled equivalently to the central impurity spin. As mentioned in \sref{sec:symm}, our model \eref{eq:heff} in this limit has the same NFL fixed point as the 4CK model, and is therefore expected to show similar low-energy physics.\cite{Fabrizio1994,Fabrizio1996,Sengupta1996} This is confirmed below. We also look beyond the low-energy regime, calculating exactly with NRG the full universal crossover behavior of the thermodynamics and dynamics.


\subsection{Thermodynamics}
\label{sec:thermiso}

\begin{figure}
\includegraphics[width=8.5cm]{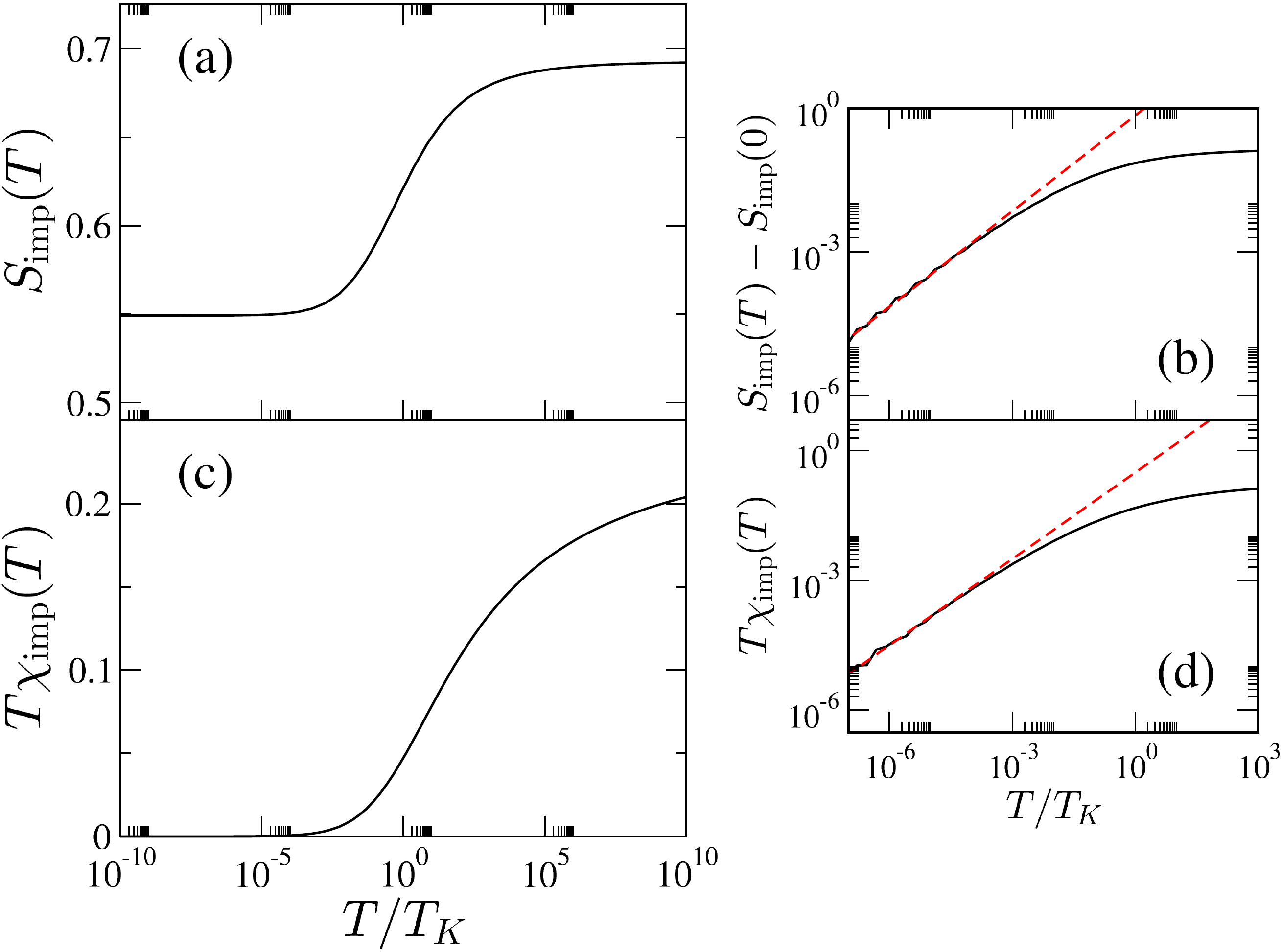}
\caption{\label{fig:td_uni}Universal thermodynamics in the channel-isotropic limit. (a) The `impurity entropy' $S_\mathrm{imp}(T)$ showing the crossover from the local moment fixed point value of $\ln 2$ when $T\gg T_K$, to the value $\ln\sqrt{3}$ at the NFL fixed point as $T\to 0$. (b) The leading $T$ dependence of $S_\mathrm{imp}(T) - S_\mathrm{imp}(0)$ from NRG (solid line), compared to the asymptotic form $\sim(T/T_K)^{2/3}$ (dashed line). (c) The impurity spin susceptibility $T \chi_\mathrm{imp}(T)$ showing the same crossover as in (a). (d) The leading $T$ dependence of $T\chi_\mathrm{imp}(T)$ from NRG (solid line), again compared to the asymptotic form $\sim(T/T_K)^{2/3}$ (dashed line).}
\end{figure}

The RG flow associated with the isotropic model is seen most vividly in the `impurity entropy' as a function of temperature, $S_\mathrm{imp}(T)$, defined as the difference in entropy between $\hat{H}$ and $\hat{H}_\mathrm{leads}$. The Kondo temperature, $T_K$, sets the scale for universal flow from the LM fixed point to the NFL fixed point. In practice, we define it numerically via $S_\mathrm{imp}(T_K) = \tfrac{1}{2}(\ln \sqrt{3} + \ln 2) \simeq 0.62$, suitably halfway between the limiting fixed point entropies. Numerical results for any choice of the bare parameter $\rho J$ then collapse onto a single universal scaling curve when plotted in terms of $T/T_K$. This scaling curve is shown in \fref{fig:td_uni}(a).

The general form of the entropy scaling curve for the model \eref{eq:heff} in the isotropic limit is indeed as one would expect for the 4CK model: a crossover on the scale of $T_K$ from the LM fixed point with $S_\mathrm{imp}(T)=\ln 2$ to the NFL fixed point, with residual $T=0$ entropy\cite{Andrei1984,*Tsvelik1985,affleck1990current} $S_\mathrm{imp} = \ln\sqrt{3} \simeq 0.55$ (this non-trivial value is reproduced very accurately in our NRG calculations). Behavior in the vicinity of the fixed point is characteristic of the non-Fermi-liquid physics, with leading corrections in the low-temperature limit $(T/T_K)\ll 1$ of $S_\mathrm{imp}(T) - S_\mathrm{imp}(0) \sim (T/T_K)^{2/3}$, as shown in \fref{fig:td_uni}(b) by comparison to the dashed line. This behavior arises in the 4CK model\cite{Andrei1984,*Tsvelik1985,affleck1990current} due to a leading irrelevant operator of scaling dimension $4/3$. The leading irrelevant operator at the NFL fixed point of \eref{eq:model} was identified in Ref.~\onlinecite{Beri2012}, and has the same scaling dimension.

\Fref{fig:td_uni}(c) shows a similar universal plot of the impurity spin susceptibility $T\chi_\mathrm{imp}(T)$ vs $T/T_K$. The low $T/T_K\ll 1$ behavior [see \Fref{fig:td_uni}(d)] is found to be $T\chi_\mathrm{imp}\sim(T/T_\mathrm{K})^{2/3}$, again entirely consistent with 4CK-like physics.\cite{Andrei1984,*Tsvelik1985,affleck1990current}

\begin{figure}
\includegraphics[width=7.5cm]{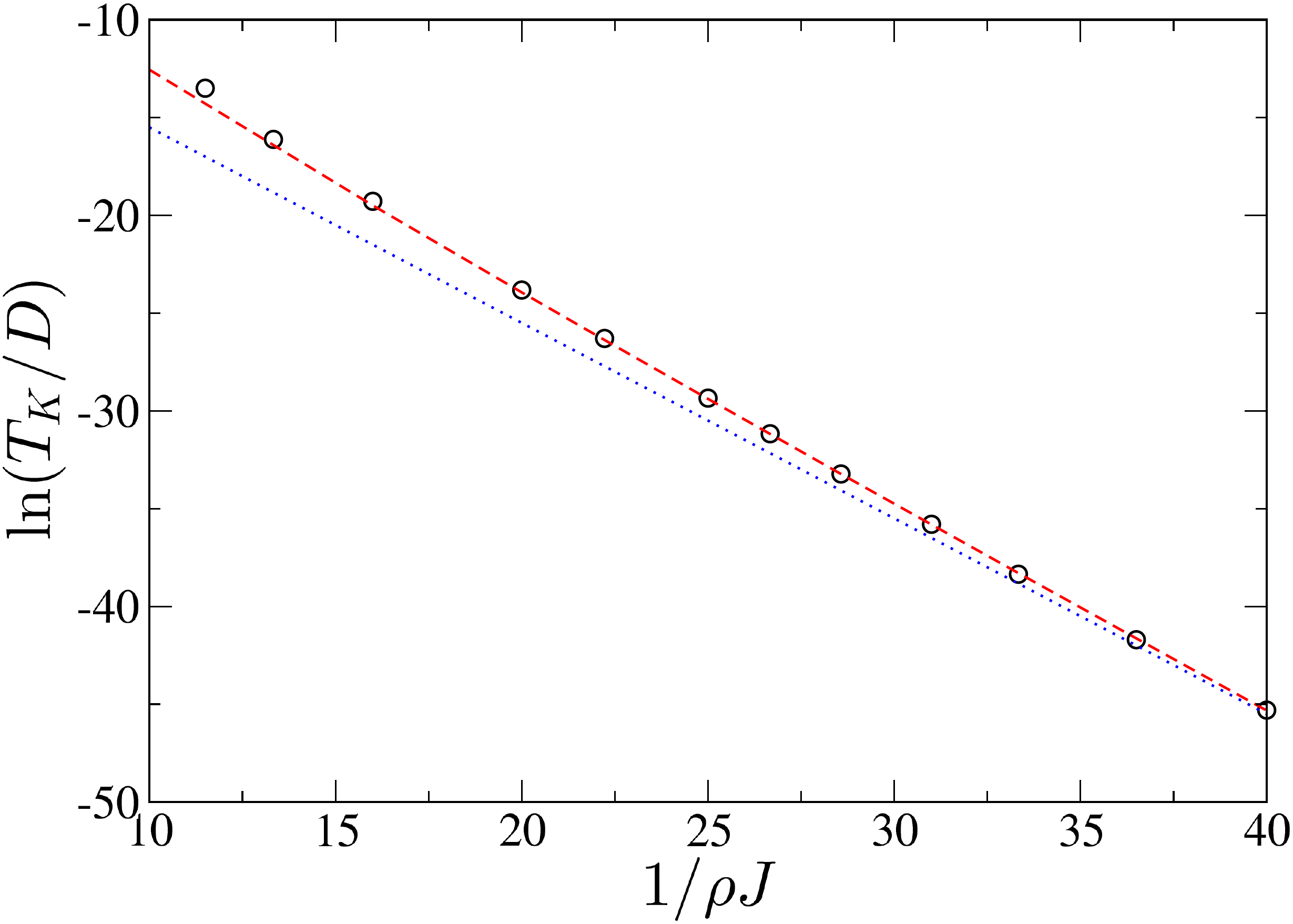}
\caption{\label{fig:tkiso} The dependence of the Kondo scale $T_K$ on $1/(\rho J)$ in the channel-isotropic limit. Circles are NRG data, the dashed line is the form given in \eref{eq:tk}, and the blue dotted line is the asymptotic form $\ln(T_K/D)\sim -1/(\rho J) + c'$.}
\end{figure}


\subsection{Kondo temperature}
\label{sec:tkiso}
Having examined the universal thermodynamics as a function of $T/T_K$, we now consider the dependence of the Kondo temperature, $T_K$, on the coupling strength, $J$. The asymptotic behaviour for small $\rho J$ can be obtained by applying perturbative scaling\cite{Anderson1970} to \eref{eq:heff}, which yields
\begin{equation}
\label{eq:tk}
T_K/D = c(\rho J)^2 \exp[-1/(\rho J)] \;,
\end{equation}
to third-order. This result is tested and confirmed in \fref{fig:tkiso}, where we plot $\ln(T_K/D)$ vs $1/\rho J$ [points obtained by NRG, dashed line is \eref{eq:tk}]. It is worth pointing out here that \eref{eq:tk} is also obtained for the 4CK model from perturbative scaling, suggesting that the RG flow from the (same) LM to NFL fixed points in the two models are rather similar, at least for small $\rho J$.

For comparison, the \emph{second-order} perturbative scaling result, $T_K/D = c' \exp[-1/(\rho J)]$, is plotted as the dotted line in \fref{fig:tkiso}. At this level one does not obtain the prefactor $(\rho J)^2$ in front of the exponential in \eref{eq:tk}; from the figure it is clear that this prefactor has a rather strong influence on the Kondo scale for moderate values of $\rho J$.


\subsection{Scattering t-matrix}
\label{sec:tmatrix}
We turn now to the dynamics of the model \eref{eq:heff}, about which far less is currently known. Asymptotic behavior near the NFL fixed point of the 4CK model has been extracted using CFT,\cite{affleck1990current,*affleck1991kondo,*affleck1991critical,*Affleck1993} but full crossover functions for either model have not previously been calculated. 

The scattering t matrix is a central quantity of interest, which contains rich information about the RG flow and underlying physics. It describes scattering between eigenstates of the disconnected leads, induced by the `impurity', and is thus defined for a given channel, $m$, in terms of the retarded Green function for conduction electrons in that channel,
\begin{equation}
\label{eq:gkkdef}
G_{k k',m}(t) = -\I\theta(t)\langle\{\cdes{k m}(t), \ccre{k' m}(0)\}\rangle.
\end{equation} 
Equations of motion\cite{Zubarev1960} for the Fourier transformed Green function, $G_{kk',m}(\omega)$, then yield directly
\begin{equation}
G_{k k',m}(\omega) = G^0_{k k',m}(\omega)  +  G^0_{k k,m}(\omega) \tau_{kk',m}(\omega) G^0_{k' k',m}(\omega)
\end{equation} 
where $G^0_{k k',m}(\omega)=\delta_{k k'}/(\omega+ \I 0^+ -\epsilon_{k})$ is the Green function of the free leads in the absence of the impurity. $\tau_{kk',m}(\omega)$ is the t matrix, which contains information about electronic correlations and the Kondo effect. We consider its spectrum,
\begin{equation}
\label{eq:tnudef}
t_{m}(\omega) = - \pi \rho\, \mathrm{Im} \sum_{k}\tau_{kk,m}(\omega) \;.
\end{equation}

Note that, although $t_m(\omega)$ is defined in terms of the lead operators in the rotated basis of \eref{eq:heff}, it also describes scattering in the physical basis of \eref{eq:model}. This follows because the transformation \eref{eq:3trans} is trivial for $m=0$ so that, from Eqns.~(\ref{eq:gkkdef})--(\ref{eq:tnudef}), the t matrix of the physical lead $j=3$ is simply given by that for channel $m=0$ in the rotated basis. Then, channel isotropy implies that the t matrices for all physical and rotated channels are identical. In practice we calculate the single t matrix from $t_{m=0}(\omega)$ with NRG. Details can be found in Appendix~\ref{sec:appt}.

\begin{figure}
\includegraphics[width=8.5cm]{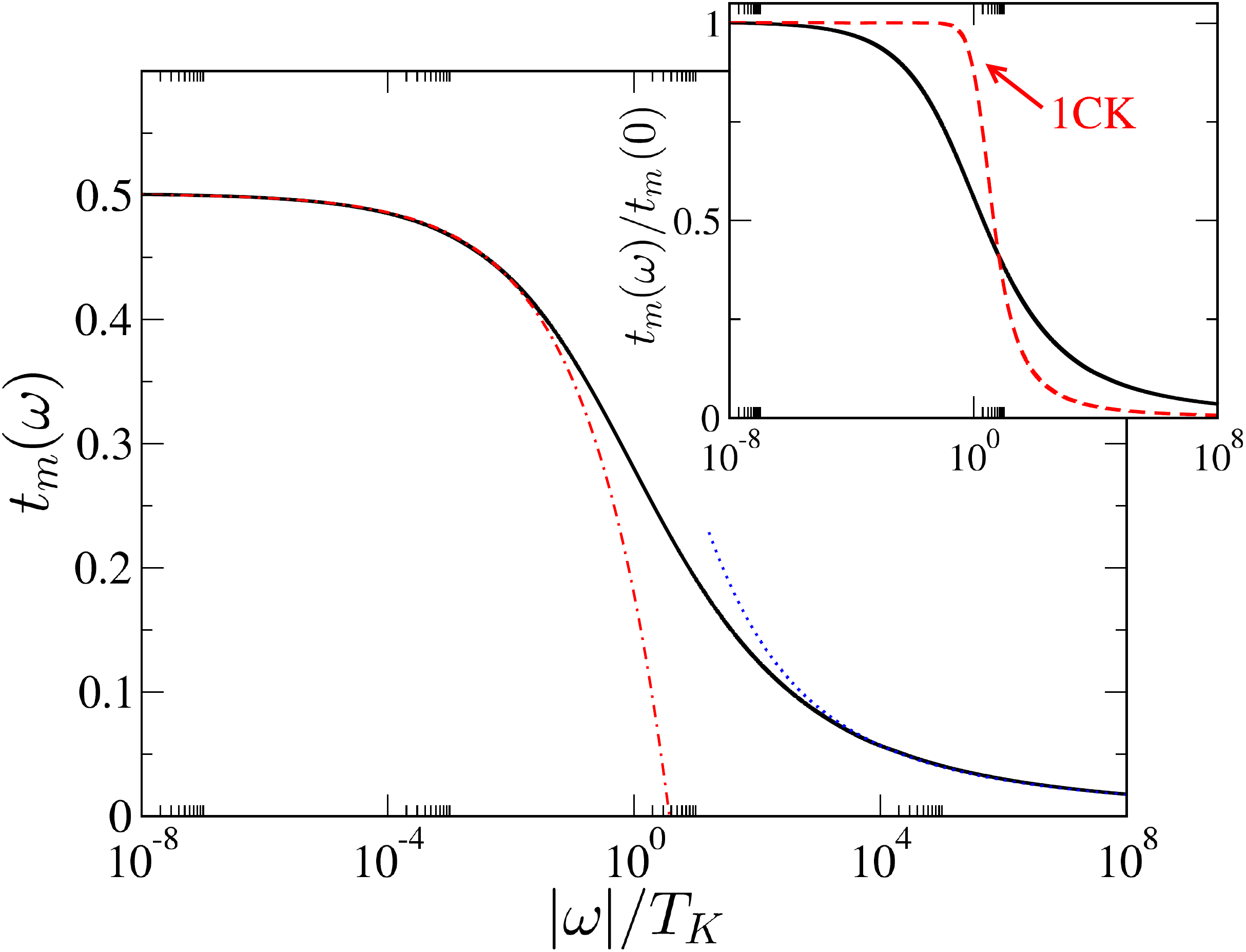}
\caption{\label{fig:tmatrix} The universal form of the scattering t-matrix $t_m(\omega)$ in the channel isotropic limit. Main figure: the solid line shows $t_m(\omega)$ calculated from NRG, as well as the low-frequency asymptotic form $\sim (|\omega|/T_K)^{1/3}$ (dot-dashed line) and high-frequency logarithmic tail (dotted line) discussed in the text. The inset compares $t_m(\omega)/t_m(0)$ for the present model (solid line) to that of the one-channel Kondo model (dashed line).}
\end{figure}

The universal RG flow between the LM and NFL fixed points is naturally reflected in the $T=0$ spectrum $t_m(\omega)$, the scaling curve for which we show in \fref{fig:tmatrix} in terms of $|\omega|/T_K$ (solid line obtained by NRG). 
The full crossover can be understood in terms two asymptotic limits. The high-frequency `tail' in $t_m(\omega)$, obtained for $|\omega|/T_K \gg 1$, is associated with simple perturbative scattering of conduction electrons from the spin-$\tfrac{1}{2}$ impurity local moment. The resulting behavior of the spectrum (which is common to all problems in which local moment physics plays a key role), takes the form\cite{Dickens2001}
\begin{equation}
\label{eq:logtailsform}
t_m(\omega) \overset{|\omega|\gg T_K}{\sim} \frac{1}{a + \ln^2(b~|\omega|/T_K)} \;,
\end{equation}
with $a$ and $b$ constants. A fit to this form is shown as the dotted line in \fref{fig:tmatrix}, which  our exact numerics approach asymptotically for $|\omega|/T_K \gg 1$. In the opposite regime $|\omega|/T_K\ll 1$, the t matrix has the characteristic behavior
\begin{eqnarray}
\label{eq:tmatrixasym}
t_m(\omega) - t_m(0) \overset{|\omega|\ll T_K}{\sim} - (|\omega|/T_K)^{1/3} + \mathcal{O} [(|\omega|/T_K)^{2/3} ] \;, \;\;\;\;\;\;\;\;
\end{eqnarray}
with $t_m(0)=\tfrac{1}{2}$. This power-law form is as predicted by CFT: generally one expects $t_m(\omega) - t_m(0) \sim (|\omega|/T_K)^{x-1}$, with $x$ the scaling dimension of the leading irrelevant operator near the stable fixed point; here, $x=4/3$,\cite{Beri2012} thus yielding \eref{eq:tmatrixasym}. Excellent agreement between \eref{eq:tmatrixasym} (dot-dashed line in \fref{fig:tmatrix}) and the exact NRG result is found over a wide range of energies. 

As pointed out earlier, the $t$-matrix exemplifies certain key differences between the low energy physics of the 4CK and the present model. 
This is seen immediately in the zero-frequency values: for the 4CK model, CFT predicts\cite{affleck1990current} $t^{4CK}_{\alpha}(0)=(3-\sqrt{3})/6$, while here it recovers the value $t_m(0) = \tfrac{1}{2}$ observed in \fref{fig:tmatrix}. This simply reflects the different forms of the conduction channels of the two models (four spin-$\tfrac{1}{2}$ channels versus one spin-$1$ channel); indeed, the result $t_m(0) = \tfrac{1}{2}$ is correctly recovered using the CFT fusion rules of Ref.~\onlinecite{affleck1990current} with spin-$1$ conduction electrons.

The inset of \fref{fig:tmatrix} compares the universal scaling spectrum of the t matrix for \eref{eq:heff} (solid line) to that known for the regular one-channel Kondo (1CK) model (dashed line), both plotted as $t_{m}(\omega)/t_{m}(0)$ for ease of comparison. 
\footnote{For the 1CK model, $t^{1CK}_{m}(0)=1$; and we define $T_K$ via $S_\mathrm{imp}(T_K) = \tfrac{1}{2}\ln 2$.} 
The two curves are strikingly different: the crossover from the high-frequency tails to the low-frequency power-law is much sharper in the 1CK case than for the present model. This is consistent with the very different leading corrections to the stable fixed points of the two models. Fermi-liquid theory for the 1CK model predicts\cite{Hewson1997} the ubiquitous quadratic approach to the Fermi level value, $t^{1CK}_{m}(\omega)-t^{1CK}_{m}(0)\sim (\omega/T_K)^2$, as compared to the slower NFL corrections of \eref{eq:tmatrixasym}.


\subsection{Linear conductance}
\label{sec:isocond}
Finally, we consider the zero-bias differential conductance --- the central quantity of experimental relevance: measuring this simply amounts to measuring a current in one of the leads as a response to infinitesimally small voltages. 

We employ the Kubo formalism\cite{Izumida1997} to obtain an expression amenable to treatment within NRG (see Appendix~\ref{sec:appk} for details). Our focus is the conductance in an AC field. As such, we assume that the system is in equilibrium at time $t=-\infty$, with all three leads at a common chemical potential $\mu = 0$. The chemical potential of lead $\beta$ is then given a time dependence $\mu_\beta(t) = eV_\beta \cos(\omega t)$, switched on adiabatically from $t=-\infty$. Denoting the total electronic number operator of lead $\alpha$ by $\hat N_\alpha$, the current $I_\alpha(t; \omega) = e\langle \tfrac{\D}{\D t}\hat N_\alpha\rangle$ flowing into lead $\alpha$ has reached an oscillatory steady state by time $t=0$. The dimensionless AC conductance tensor is then defined as
\begin{equation}
\left(\frac{e^2}{h}\right)G_{\alpha\beta}(\omega; T) = \lim_{V_\beta \to 0} \frac{\partial I_\alpha(t=0;\omega)}{\partial V_\beta}
\end{equation}
which holds at any temperature $T$. On taking the $\omega\to 0$ limit one obtains the static conductance,
\begin{equation}
\label{eq:cond_dc}
G^\mathrm{DC}_{\alpha\beta}(T) = \lim_{\omega \to 0} ~G_{\alpha\beta}(\omega; T) \;,
\end{equation}
in response to a DC voltage $\mu_\beta(t) = eV_\beta$ switched on adiabatically from $t=-\infty$.

In the channel-isotropic limit under consideration in this section, the off-diagonal elements $G_{\alpha\beta}(\omega)$ for $\alpha\ne\beta$ are identical by symmetry. Moreover, as shown in Appendix~\ref{sec:appk}, they are related to the diagonal elements by
\begin{equation}
G_{\alpha\beta}(\omega;T) = - \tfrac{1}{2}G_{\beta\beta}(\omega;T) \;.
\end{equation}
Physically, the minus sign reflects the fact that a positive bias applied to lead $\beta$ produces net current flow from lead $\beta$ to leads $\alpha\ne\beta$ --- i.e., positive current towards lead $\alpha$ and negative current towards lead $\beta$. The entire conductance tensor is thus fully determined by a single element; we calculate $G_{10}(\omega;T)$ explicitly below.

\begin{figure}
\includegraphics[width=8.5cm]{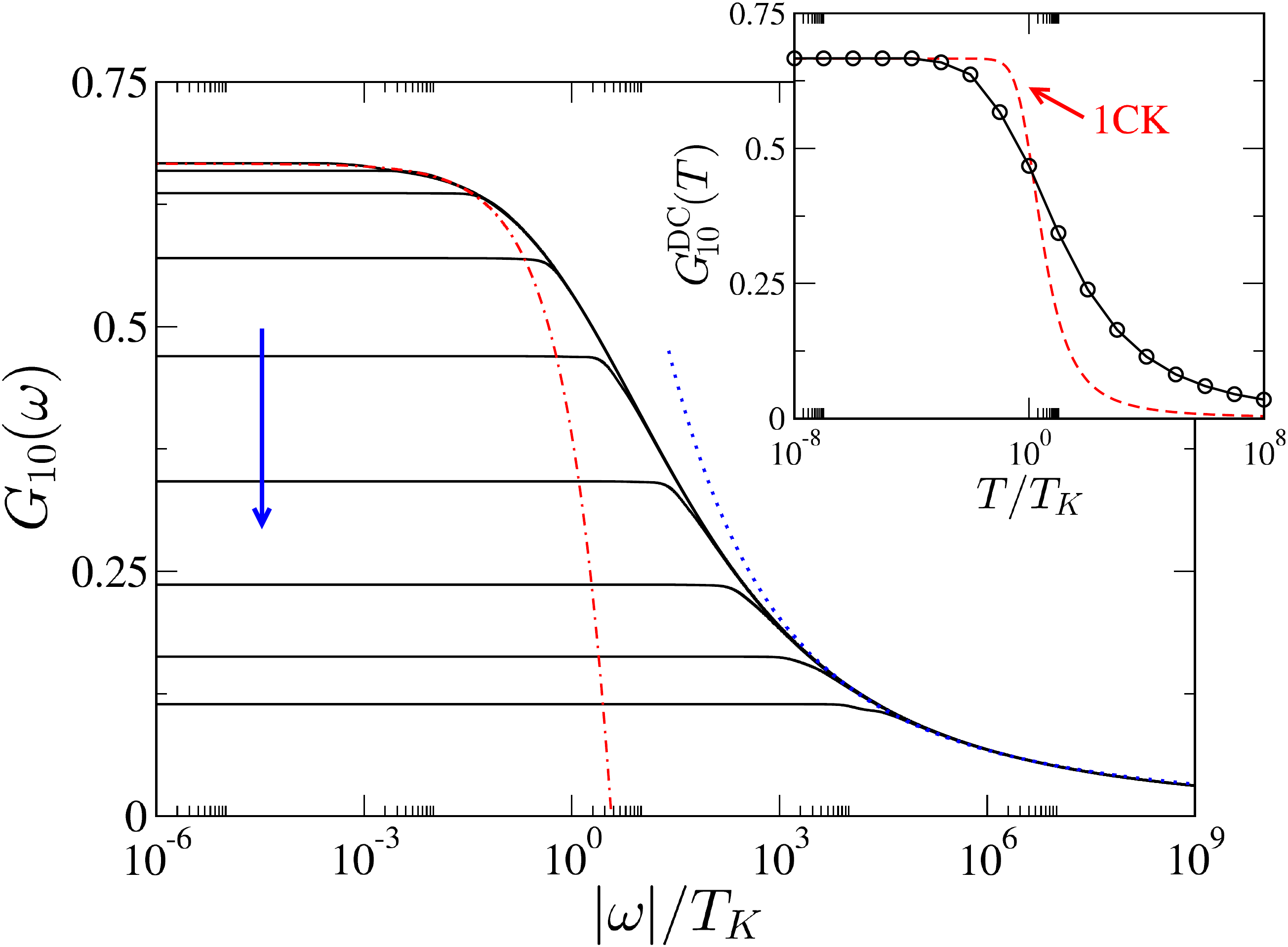}
\caption{\label{fig:cond} Universal linear conductance of the model in the channel-isotropic limit. The main figure shows the AC conductance $G_{10}(\omega)$ for $T=0$,  and for a range of temperatures $T/T_K = 10^x$ with integral $x = -4 \to +4$ (solid lines, in the direction of the arrow). The low frequency asymptote $\sim(|\omega|/T_K)^{2/3}$ (red dot-dashed line) and the high frequency logarithmic tail (blue dotted line) are also shown. Inset: the temperature dependence of the DC conductance $G_{10}^\mathrm{DC}(T)$ (circles, solid line is a guide to the eye), compared to the DC conductance of the one-channel Kondo model (dashed line). Note that we have rescaled the height of the latter by a factor of $1/3$ to facilitate comparison between the curves.}
\end{figure}

As one would expect from the preceding results, we find from NRG that $G_{10}(\omega)$ is a universal function of $\omega/T_K$ at any given temperature $T/T_K$. The main panel of \fref{fig:cond} shows the scaling form of $G_{10}(\omega)$ vs $|\omega|/T_K$ both at zero temperature, and for a range of temperatures $T/T_K = 10^x$ with integral $x = -4 \to +4$, increasing in the direction of the arrow. 

As with the t matrix, the form of $G_{10}(\omega)$ at $T=0$ can be understood in terms of its $|\omega| \lessgtr T_K$ asymptotes. At high frequencies, one observes conductance signatures characteristic of spin-flip scattering of conduction electrons from an asymptotically-free impurity spin, again of the form
\begin{equation}
\label{eq:condsymhi}
G_{10}(\omega;T=0) \overset{|\omega|\gg T_K}{\sim}  \frac{1}{a' + \ln^2(b'~|\omega|/T_K)}\;,
\end{equation}
as confirmed by comparison to the dotted line in the main panel of \fref{fig:cond}. On the scale of $|\omega|\sim T_K$, RG flow toward the NFL fixed point results in an enhancement of the conductance. At low frequencies $|\omega|\ll T_K$, corrections to the NFL fixed point yield characteristic power-law behavior,
\begin{equation}
\label{eq:condsym}
G_{10}(\omega;T=0)- G_{10}(0,0)  \overset{|\omega|\ll T_K}{\sim} -(|\omega|/T_K)^{2/3} \;,
\end{equation}
with $G_{10}(0,0)=\tfrac{2}{3}$. NRG results fold onto \eref{eq:condsym} for $|\omega|\ll T_K$, as shown by comparison with the dot-dashed line in \fref{fig:cond}.
\footnote{NRG recovers the exact value $G_{10}(0,0)=\tfrac{2}{3}$ to within a few percent (although numerical errors can in principle be systematically reduced). In our analysis of the low-frequency power-law behavior of  $G_{10}(\omega,0)$, we therefore treated $G_{10}(0,0)$ as a fitting parameter. A fit to the form $(|\omega|/T_K)^{2/3}$ extends to the lowest frequencies, compared with other simple rational powers.}  
We note that the leading correction here is of the form $(|\omega|/T_K)^{2/3}$ --- different from the $(|\omega|/T_K)^{1/3}$ behavior observed for the t matrix, see \eref{eq:tmatrixasym}. This difference is understood by a simple extension of the CFT arguments given in Ref.~\onlinecite{Beri2012}: perturbation theory around the NFL fixed point in the leading irrelevant operator yields corrections to the conductance which vanish to first order, by symmetry. The expected $(|\omega|/T_K)^{1/3}$ behavior is thus replaced by leading second-order corrections of the form in \eref{eq:condsym}.

The same basic crossover is observed in the static DC conductance, $G_{10}^\mathrm{DC}(T)$, as a function of temperature $T/T_K$. This is shown in the inset to \fref{fig:cond} as the circle points, obtained from the calculated\footnote{
We use the standard broadening scheme\cite{Bulla2008,Weichselbaum2007,*Peters2006} for calculating NRG spectra on temperature scales $T\lesssim \omega$, which has been shown\cite{Mitchell2012a} to reproduce accurately the full frequency and temperature dependence of dynamical quantities for other quantum impurity problems where exact results are known
}
$G_{10}(\omega;T)$ as $\omega\to 0$, according to \eref{eq:cond_dc} (the solid line is a guide to the eye). 

We find that the low-temperature behavior of $G_{10}^\mathrm{DC}(T)$ in the inset of \fref{fig:cond} is consistent with the CFT prediction of Ref.~\onlinecite{Beri2012}, 
\begin{equation}
\label{eq:condsymT}
G_{10}^\mathrm{DC}(T=0) - G_{10}^\mathrm{DC}(T) \overset{|\omega|\ll T_K}{\sim} -(T/T_K)^{2/3} \;,
\end{equation}
mirroring the low-$\omega$ behavior of $G_{10}(\omega,T=0)$ [with $G_{10}^\mathrm{DC}(0)\equiv G_{10}(0,0)=\tfrac{2}{3}$ as before]. For comparison we also show the universal conductance of the 1CK model (dashed line). As for the t matrix in \fref{fig:tmatrix}, the crossover in the conductance of the present model is much less rapid than for the 1CK model, and should prove useful as a means of identifying the topological Kondo effect experimentally.

On that note, we end this section with a further comment. While Eqs.~(\ref{eq:condsymhi}) and (\ref{eq:condsym}) do describe the behavior of the full conductance crossover at high and low energies, in fact they do so only at \emph{very} high and low energies, respectively. The power-law form of \eref{eq:condsym} is seen only when $|\omega|/T_K\ll 10^{-2}$; and the log-tails of \eref{eq:condsymhi} are approached only for $|\omega|/T_K\gg 10^{3}$. Depending on the value of the Kondo temperature, $T_K$, in a real experiment, it seems unlikely that either or both of these asymptotes could be robustly observed, given experimental limitations (or the presence of non-universal effects and perturbations --- see also \sref{sec:breakab}). Any positive identification of the precise nature of the topological Kondo effect is therefore likely to require a more detailed comparison of the experimental conductances with the full crossover curves from NRG. (However, its {\it existence} can be demonstrated through the qualitative signature pointed out in Ref.~\onlinecite{Beri2012}: by removing any one of the three leads, all signs of the Kondo effect, including the upturn of the conductance, should disappear.)


\section{NRG results in the anisotropic case}
\label{sec:axial}
The above discussion has focussed on the isotropic limit of the model. In this section we consider the effect of breaking full SU(2) spin symmetry, introducing anisotropy of the form $J^a_{\perp} =J_{\perp}^b \ne J_z$ in \eref{eq:heff} --- this corresponds physically to changing the Majorana-lead coupling in one of the channels. The model still possesses an \emph{axial} symmetry in this case, although the same fundamental physics described below is expected even when the couplings in all three channels are distinct, as pointed out in Ref.~\onlinecite{Beri2012}. Indeed, preliminary NRG calculations in the case of full anisotropy, do indicate that the physics is robust.

As mentioned in \sref{sec:symm}, the NFL fixed point is not destabilized by lowering the symmetry in this way (the spin anisotropy is RG irrelevant, meaning that the NFL fixed point itself is isotropic).\cite{Beri2012} The stability of the NFL fixed point does not itself rule out a quantum phase transition\cite{Schiller2008} on tuning the ratio $J_z/J_{\perp}$; however, as shown below, we do find the low-temperature fixed point in all cases to be the NFL fixed point and hence the physics of the model to be robust to breaking the channel isotropy. We first address this point in more detail, before moving on to discussing further predictions on experimental aspects.


\subsection{Kondo temperature}
\label{sec:axialscale}

\begin{figure*}
\includegraphics[width=17cm]{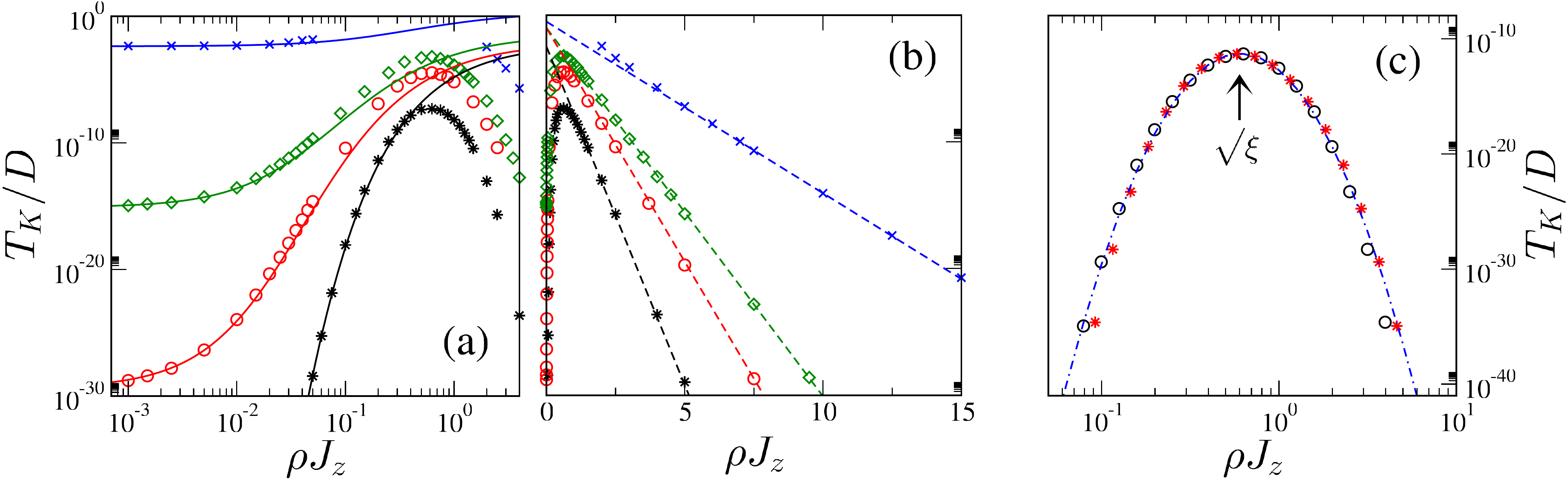}
\caption{\label{fig:tk_aniso} $\rho J_z$ dependence of the Kondo temperature $T_K/D$, in the channel anisotropic case. (a) Log-log plot, for $J_\perp/D = 0.5$, $0.1$, $0.05$ and $0.01$ (crosses, diamonds, circles and asterisks, respectively), compared to the result of perturbative scaling [solid lines, \eref{eq:tkaniso}]. (b) Same data on a log-linear plot, compared with \eref{eq:TklargeJz} [dashed lines]. (c) Crossover region for small $J_{\perp}/D=0.001$, highlighting the duality between $\rho J_z$ (circles) and $\xi/\rho J_z$ (stars), compared with \eref{eq:Tkdual} [dot-dashed line].}
\end{figure*}

It is instructive to examine how the Kondo temperature $T_K$ varies with the parameters of our effective model, \eref{eq:heff}, in the general axial-symmetric case. In \fref{fig:tk_aniso} we plot $T_K$ as a function of $\rho J_z$ for various fixed $\rho J_\perp > 0$.

The low $\rho J_z$ $(\ll 1)$ behavior, seen most clearly in \fref{fig:tk_aniso} (a), can be understood using perturbative scaling. As mentioned in \sref{sec:symm}, to second-order this yields scaling equations for \eref{eq:heff} that are identical to the regular spin-$\tfrac{1}{2}$ Kondo model.\cite{Beri2012} The equations can be solved\cite{Anderson1970a,Zitko2008} to give 
\begin{subequations}
\label{eq:tkanisogrp}
\begin{equation}
\label{eq:tkaniso}
T_K/D \;\; \overset{\rho J_z \ll 1}{\sim}  \;\; c\, \exp\left[-\frac{\alpha}{\rho J_z}\right] 
\end{equation}
where
\label{eq:tkalpha}
\begin{align}
\alpha &= \begin{cases} (\tan^{-1}\gamma)/\gamma \qquad &: \qquad J_z \le J_\perp \\
(\tanh^{-1}\gamma)/\gamma  \qquad &: \qquad J_z \ge J_\perp \end{cases} 
\end{align}
with
\begin{equation}
\gamma^2 = \left |1- \left(J_\perp/J_z \right)^2 \right| \;.
\end{equation}
\end{subequations}
When $\rho J_z \ll 1$, there is excellent agreement between exact NRG results for $T_K$ (points) and \eref{eq:tkaniso} [solid lines, panel (a)]. In each case, we have adjusted the constant $c$ to fit the data, implying a weak dependence on $\rho J_{\perp}$ of the pre-exponential factor in \eref{eq:tkaniso}, obtained to higher order in perturbative scaling. 

When $\rho J_z $ becomes large, the perturbative treatment naturally breaks down. As seen from NRG results in \fref{fig:tk_aniso}, the Kondo temperature in fact passes through a maximum at $\rho J_z \sim 1$, and then decreases rapidly as $\rho J_z$ is further increased. Specifically, we find from NRG for $\rho J_z \gg 1$ that,
\begin{equation}
\label{eq:TklargeJz}
T_K/D  \;\; \overset{\rho J_z \gg 1}{\sim} \;\;  \exp\left[-f(\rho J_\perp)\times \rho J_z\right]  \;,
\end{equation}
which is plotted for comparison as the dashed lines in \fref{fig:tk_aniso} (b). On extrapolating this result, we conclude that that $T_K$ remains finite for any $\rho J_z$ (provided $\rho J_{\perp}>0$). (We have also performed a direct survey of the parameter space of the model to substantiate this conclusion further.) As such, while the Kondo temperature is very small for large $\rho J_z$, there is always overscreening of the `impurity' for antiferromagnetic $\rho J_z$.

The latter point may not be surprising at first sight. However, in the case of the 4CK model (with the same low-energy NFL fixed point), the ground state is a free local moment state for a sufficiently large, antiferromagnetic $\rho J_z$.\cite{Schiller2008} The argument involves a mapping between the positive and negative $J_z$ sectors, and hence the transition for antiferromagnetic $\rho J_z$ is effectively the same Kosterlitz-Thouless transition well known to arise in the ferromagnetic case.\cite{Schiller2008} The absence of such a transition here reflects that, while the fixed points of the two models are the same, the RG flows between these fixed points are very different in the general case of exchange anisotropy (c.f. the discussion of \sref{sec:tkiso}). 

Further analytical insight is obtained by applying the abelian bosonization technique of Ref.~\onlinecite{Schiller2008} to the present problem. We predict a duality in \eref{eq:heff}:
\begin{equation}
\label{eq:duality}
\hat{H}_{\text{eff}}\left ( \rho J_{\perp},\rho J_{z} \right ) \leftrightarrow \hat{H}_{\text{eff}} \left ( \rho J_{\perp},\frac{\xi}{\rho J_{z}} \right ) \;,
\end{equation}
with $\xi = (2/\pi)^2\simeq 0.41$. This result does not hold ubiquitously, however, as the bosonization argument requires that $\rho J_\perp \ll \rho J_z$ and $\rho J_\perp \ll \xi/(\rho J_z)$. To understand the effect of these constraints, it is convenient to define the quantity
\begin{equation}
\nu = \ln (\rho J_z) - \tfrac{1}{2}\ln \xi, 
\end{equation}
from which it follows simply that \eref{eq:duality} holds when
\begin{equation}
\label{eq:boshold}
|\nu| \ll |\ln(\rho J_\perp)-\tfrac{1}{2}\ln\xi|.
\end{equation}
and thus becomes most valid when $|\nu| \to 0$, i.e. $\rho J_z \sim \sqrt\xi\simeq 0.64$. 

Equation~(\ref{eq:duality}) is consistent with the absence of a transition for antiferromagnetic $J_z$ since, in contrast to the 4CK model,\cite{Schiller2008} the duality does not change the sign of the exchange coupling and therefore does not map the ferromagnetic Kosterlitz-Thouless transition of the model onto the antiferromagnetic side. Instead it simply maps small $J_z$ to large $J_z$; in this sense it is similar to the duality inherent to the two-channel Kondo (2CK) model.\cite{Kolf2007}

A further consequence of \eref{eq:duality} can be seen by writing \eref{eq:duality} in terms of $\nu$ as 
\begin{equation}
\label{eq:duality2}
\hat{H}_{\text{eff}}\left ( \rho J_{\perp},\sqrt{\xi}\mathrm{e}^\nu \right ) \leftrightarrow \hat{H}_{\text{eff}} \left (\rho J_\perp, \sqrt{\xi} \mathrm{e}^{-\nu} \right ).
\end{equation}
Since the Hamiltonian is invariant to changing the sign of $\nu$, it follows that
\begin{equation}
\label{eq:tkdual}
T_K(\nu) = T_K(-\nu)
\end{equation}
for fixed $\rho J_\perp$ when \eref{eq:boshold} is satisfied. On making the ansatz that $\ln T_K$ has a power series expansion in $\nu$, one thus obtains 
 \begin{align}
T_K/D \;\;  \overset{\rho J_z \sim \sqrt{\xi}}{\sim}& \;\;  \exp(- u \nu^2)\\
&=\exp\left\{ -u\,\left[\ln(\rho J_z) - \tfrac{1}{2}\ln(\xi)\right]^2 \right\} 
\label{eq:Tkdual}
\end{align}
with $u$ a constant: this form agrees very well with our NRG results when \eref{eq:boshold} is satisfied, albeit with a slightly adjusted $\xi \simeq 0.37$. A fit of \eref{eq:Tkdual} to NRG data for $\rho J_\perp = 0.001$ is shown in \fref{fig:tk_aniso} (c) as the dot-dashed line.

The duality $\rho J_z \leftrightarrow \xi/(\rho J_z)$ also sheds light on \eref{eq:TklargeJz}. For $J_\perp \ll J_z \ll 1$, the perturbative scaling result of \eref{eq:tkanisogrp} is valid, and can be expanded to give
\begin{align}
\alpha &\sim - \ln(J_\perp/J_z) + \mathcal{O}[(J_\perp/J_z) \ln(J_\perp/J_z)]\\
&\sim -\ln(\rho J_\perp)
\end{align}
for $\ln(\rho J_\perp) \ll \ln(\rho J_z)$, 
such that
\begin{equation}
T_K/D \sim v \exp\left[\ln(\rho J_\perp)/\rho J_z\right],
\end{equation}
with $v$ a constant. 
But since $\ln(\rho J_\perp) \ll \ln(\rho J_z)$ it follows from \eref{eq:boshold} that the duality of \eref{eq:duality} holds, and hence 
\begin{equation}
T_K/D \sim v \exp\left[\frac{1}{\xi}\ln(\rho J_\perp) \rho J_z\right],
\end{equation}
for $\rho J_z \gg 1$. This recovers \eref{eq:TklargeJz}, with $f(\rho J_\perp) = -\ln(\rho J_\perp)/\xi$ asymptotically for small $\rho J_\perp$.


\subsection{Physical quantities}
\label{sec:axialdyn}

In addition to the results above, we have calculated the t matrices when $J_\perp \ne J_z$. On lowering the channel symmetry we find as expected that $t_1(\omega)=t_{-1}(\omega) \ne t_0(\omega)$. Specifically, the high-frequency logarithmic tails of the t matrices still take the form of \eref{eq:logtailsform} but with different constants. At frequencies $|\omega|\lesssim T_K$, however, we find that the t matrices fall onto the same universal curve as in \fref{fig:tmatrix}. This reflects an emergent channel symmetry at the NFL fixed point.\cite{Beri2012}

Similar behavior is naturally found when examining the AC conductance. We thus conclude that channel isotropy would not be required in order to observe the non-Fermi-liquid physics of the model described in \sref{sec:resiso}.


\section{Fermi-liquid crossover}
\label{sec:breakab}

Finally, we consider breaking the axial symmetry of \eref{eq:heff} by taking $J_\perp^a\ne J_\perp^b$. As shown in \sref{sec:modelderiv}, this kind of symmetry breaking arises due to non-local Majorana-Majorana, Majorana-lead, or lead-lead couplings in the bare model, \eref{eq:Htun}. In \sref{sec:symm} we argued that such perturbations are RG relevant, and drive the system to a strong coupling, Fermi-liquid (FL) ground state. In this section we confirm this explicitly using NRG. Although even tiny perturbations destabilize the  NFL fixed point, NFL physics may still be observable at finite energies/temperatures.

\begin{figure}
\includegraphics[width=8.5cm]{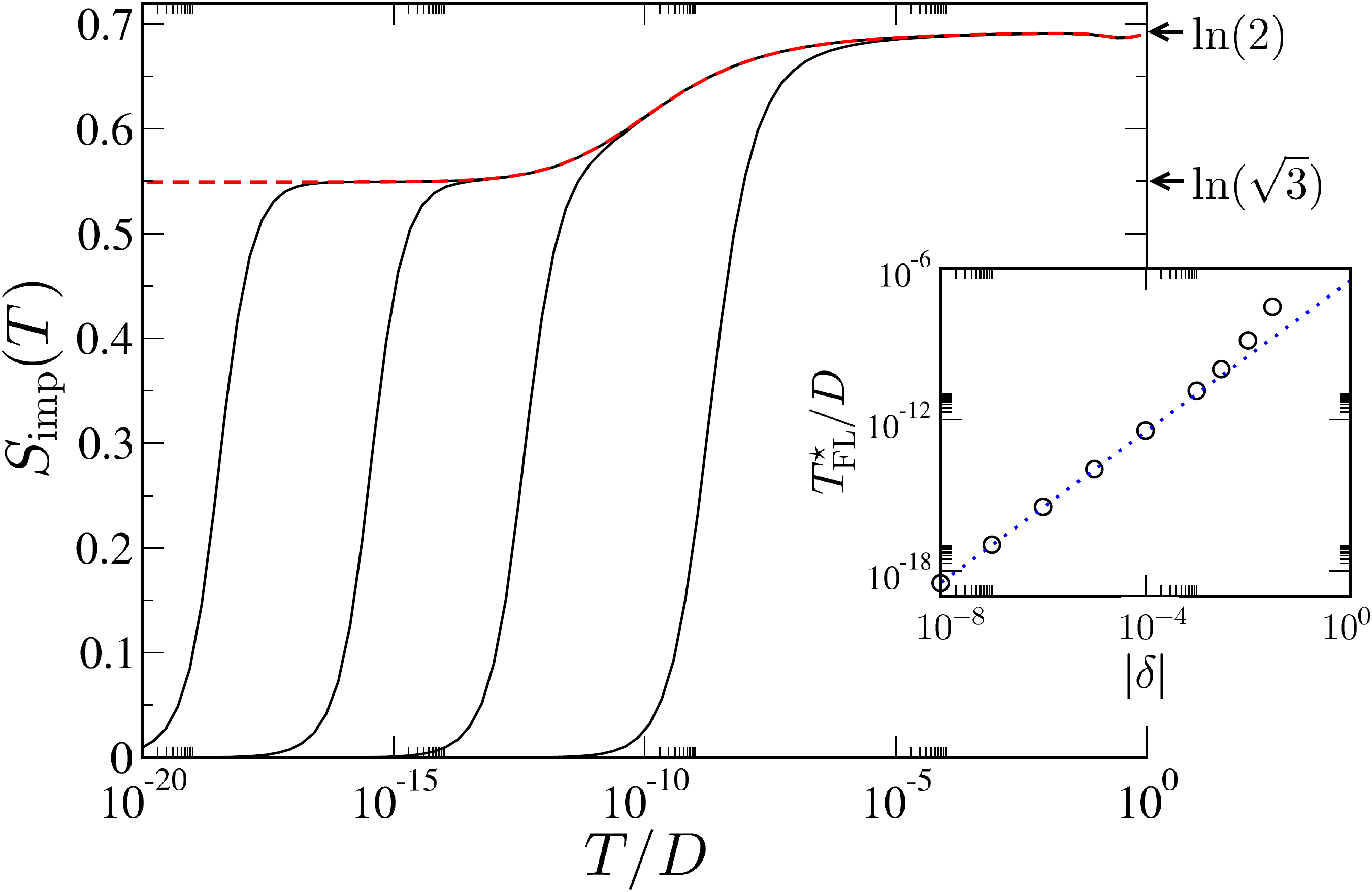}
\caption{\label{fig:S_pert} Impurity entropy in the case $J_{\perp}^a\ne J_{\perp}^b$, resulting from cross-couplings in \eref{eq:pert}. The main figure shows $S_\mathrm{imp}(T)$ for $J_\perp^a/D = 0.1 + \delta$, $J_\perp^b/D = J_z/D = 0.1$, with $\delta = 10^{-2}$, $10^{-4}$, $10^{-6}$ and $10^{-8}$ (solid lines), together with the $\delta = 0$ result (dashed line). For $\delta\ne 0$ the entropy crosses over to the strong coupling value $S_\mathrm{imp}=0$ on the scale $T_\mathrm{FL}^*$. The inset shows the dependence of $T_\mathrm{FL}^*$ on $\delta$ (circle points), compared to the asymptotic form of \eref{eq:tflstar} (dotted line).}
\end{figure}

In \fref{fig:S_pert} we show NRG results for the impurity entropy, with  $J_\perp^b/D = J_z/D = 0.1$, $J_\perp^a/D = 0.1 + \delta$, and several different values of the symmetry-breaking parameter $\delta$ (solid lines), compared with the $\delta = 0$ case (dashed line). For all finite $|\delta|>0$, the $T\rightarrow 0$ entropy always vanishes, $S_\mathrm{imp}=0$. This is indicative of RG flow to the stable SC Fermi-liquid fixed point. We denote the scale characterizing this flow by $T_\mathrm{FL}^*$, defining it in practice by $S_\mathrm{imp}(T_\mathrm{FL}^*) = \tfrac{1}{4}\ln 3$, suitably halfway between the SC and NFL fixed point values. 

CFT allows one to identify relevant operators in the channel sector which have the same symmetry as the $\delta = J_{\perp}^a/D -J_{\perp}^b/D$ perturbation. Owing to axial symmetry, there is one such operator at the NFL fixed point with scaling dimension $x=1/3$. This dimension implies that the Fermi-liquid crossover scale has a power-law dependence,
\begin{equation}
\label{eq:tflstar}
T_\mathrm{FL}^* \sim T_K \times |\delta|^{3/2} \;,
\end{equation}
which should apply when $\delta$ acts as a perturbation to the NFL fixed point --- i.e., when there is good scale separation $T_{\text{FL}}^* \ll T_K$. Indeed, this result is confirmed in the inset to \fref{fig:S_pert}, by comparison of NRG results (points) to \eref{eq:tflstar} (dotted line).

Although any nonzero $\delta \ne 0$ destabilizes the NFL fixed point,  signatures of non-Fermi-liquid physics should still be observable in an intermediate temperature `window', provided the strength of perturbations is small (as might be expected physically, see \sref{sec:modelderiv}).  As seen from \fref{fig:S_pert}, when there is good scale separation $T_{\text{FL}}^*  \ll T_K$, thermodynamics at higher temperatures $T\gg T_{\text{FL}}^*$ are essentially indistinguishable from the case $\delta=0$ (where the NFL fixed point is stable down to $T=0$). The topological Kondo effect could thus be identifiable in this regime, even when symmetry-breaking perturbations act.  
Indeed, at lower temperatures, the subsequent Fermi-liquid crossover is wholly  characteristic of flow from the unstable NFL fixed point, and exhibits universal scaling in terms of $T/T_{\text{FL}}^*$. Thus, flow \emph{away} from the NFL fixed point due to symmetry-breaking perturbations could also be used to identify the topological Kondo effect.

\begin{figure}
\includegraphics[width=6.5cm]{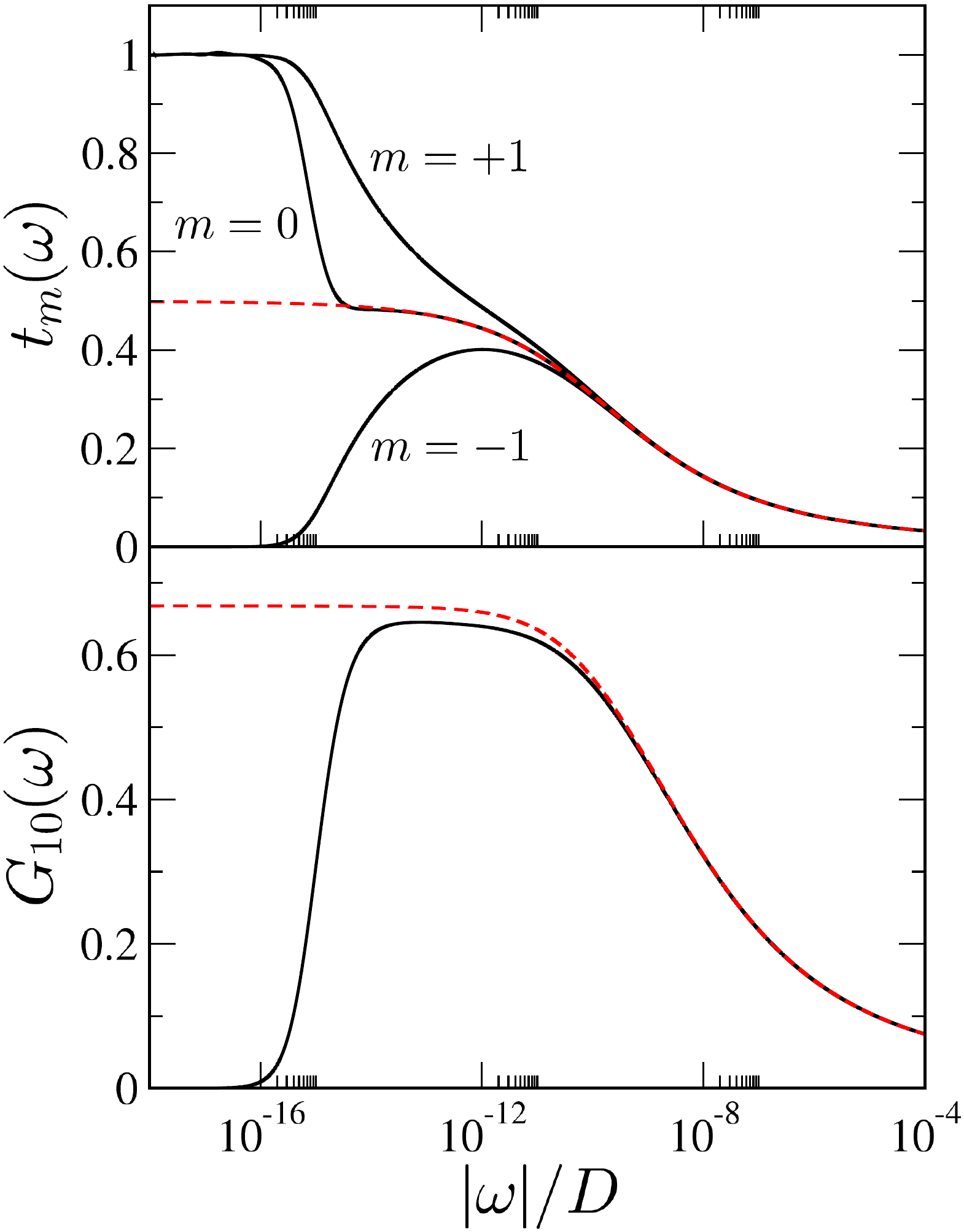}
\caption{\label{fig:dyn_anotb} Dynamics for broken $J_{\perp}^a\ne J_{\perp}^b$ symmetry, with $J_\perp^a/D = 0.1 + \delta$, $J_\perp^b/D=J_z = 0.1$ and $\delta = 10^{-6}$. The upper panel shows the three t matrices $t_m(\omega)$ (solid lines) compared with the $\delta=0$ result (dashed line). The lower panel shows the AC conductance, $G_{10}(\omega)$, for the same system.}
\end{figure}

Dynamics of the model when $J_\perp^a \ne J_\perp^b$ are of particular interest. In the top panel of \fref{fig:dyn_anotb}, we show the three t matrices $t_m(\omega)$, when $J_\perp^{a}/D = 0.1 + \delta$ and  $J_\perp^b/D = J_z/D = 0.1$, with $\delta = 10^{-6}$. The corresponding t matrices for $\delta = 0$ are shown as the dashed line (these are identical by symmetry in this isotropic limit). Here, $T_\mathrm{FL}^*$ is sufficiently small that the NFL fixed point strongly affects the RG flow and resulting t matrix lineshapes. When there is good scale separation, we find for $T_{\text{FL}}^* \ll |\omega| \ll T_K$ the asymptotic behavior of
\begin{equation}
t_m(\omega) ~~ \sim ~~ \begin{cases} \tfrac{1}{2} \pm p(|\omega|/T_{\text{FL}}^*)^{-1/3} \qquad &: \qquad m=\pm 1 \\
\tfrac{1}{2} + \mathcal{O}(|\omega|/T_\mathrm{FL}^*)^{-2/3} \qquad &: \qquad m= 0 \end{cases} \;.
\end{equation}
The behavior of $t_0(\omega)$ is consistent with first-order corrections from the relevant operator mentioned above, while the deviations of $t_{+1}(\omega)$ and $t_{-1}(\omega)$ appear to arise at the second-order level. 
Interestingly, in the basis of \eref{eq:lead_trasnform}, the t matrix in channel $m=0$ is essentially indistinguishable from the $\delta=0$ case for energies $|\omega|\gg T_{\text{FL}}^*$, and we find $t_0(\omega)=\tfrac{1}{2}[t_{-1}(\omega)+t_{+1}(\omega)]$.

At lower energies $|\omega|\ll T_{\text{FL}}^*$, the perturbation $\delta\ne 0$ grows under RG and becomes large, causing deviation from this behavior. Near the Fermi-liquid fixed point we instead find for $\delta>0$ that
\begin{equation}
t_m(\omega) ~~ \sim ~~ \begin{cases} 1 - q_m(|\omega|/T_{\text{FL}}^*)^{2} \qquad &: \qquad m=0, +1 \\
0 + q_m(|\omega|/T_{\text{FL}}^*)^{2}\qquad &: \qquad m= -1 \end{cases} \;.
\end{equation}
This is characteristic of standard strong coupling Kondo physics in leads $m=0$ and $m=+1$, with lead $m=-1$ decoupling asymptotically.
The roles of $m=\pm 1$ are reversed when $\delta<0$, since the Kondo effect then takes place with leads $m=0$ and $m=-1$. The t matrices thus indicate directly which channels are participating in the Kondo effect at the SC fixed point, and support the physical picture discussed in \sref{sec:symm}.

The low-energy crossover to the SC fixed point (and the associated `window' of non-Fermi-liquid behavior) is also seen in the AC conductance, shown in the bottom panel of \fref{fig:dyn_anotb} for the same parameters as the t matrices discussed above. We find asymptotically that
\begin{equation}
G_{10}(\omega) \sim \begin{cases} \tfrac{2}{3} - p'(|\omega|/T_{\text{FL}}^*)^{-4/3} \qquad &: \qquad T_{\text{FL}}^* \ll |\omega| \ll T_K \\
 0+ q'(|\omega|/T_{\text{FL}}^*)^{2} \qquad &: \qquad |\omega| \ll T_{\text{FL}}^* \ll T_K \end{cases}\;.
\end{equation}
Interestingly, $G_{10}\to 0$ as $|\omega|\to 0$ here, despite the fact that channels $m=0$ and $+1$ form a Kondo singlet with the central impurity spin for $\delta>0$ (typically the Kondo effect results in a zero-bias \emph{enhancement} of conductance). The vanishing conductance $G_{10}(0)=0$ (between channels $m=0$ and $+1$), is in fact due to the decoupling of the third channel, $m=-1$, at the strong coupling fixed point. This can be understood by the following physical argument: an electron tunneling from lead $m=0$ to $+1$ requires (by conservation of total $S_z$) to flip the impurity spin from $\sigma=\uparrow$ to $\downarrow$. Further electronic transport is now blocked, since the impurity mediates the current between all channels, and is already in the $\sigma=\downarrow$ configuration. Only when lead $m=-1$ is coupled in can a finite conductance result: the $\sigma=\uparrow$ impurity configuration can be restored (thus `resetting' the system) by a tunneling process from $m=-1$ to $0$. 
The Fermi-liquid ground state of the model (arising from any inter-Majorana coupling) thus inevitably results in vanishing conductance; although of course signatures of criticality and non-Fermi-liquid physics may appear at finite temperatures. Naturally, similar results are obtained for $G_{00}(\omega)=-G_{10}(\omega)-G_{-10}(\omega)$.

One might ask how the results above change when the non-local perturbations are larger, e.g. of order $T_K$ or more. In this case, the NFL fixed point (characterized by fractional power-law behaviour) seen on decreasing $\omega$ or $T$ will of course not be directly observable. The system will flow away to the SC fixed point on the energy scale of the perturbations, and hence the conductance will drop to zero  before the power-law behavior is reached. But the key signature of the topological Kondo effect itself---the \emph{approach} to the NFL fixed point, characterized by the low frequency/temperature conductance peak---could still be observable for such large perturbations. This is aided by the fact that the approach to the NFL fixed point is very slow (compared to that of the conventional Kondo effect, see \fref{fig:cond}), taking place over a very wide energy range that starts several orders of magnitude above $T_K$.


\section{Conclusion}
\label{sec:concl}
We have analyzed in detail the physics of the minimal topological Kondo setup of Fig.~\ref{fig:MF_schematic}, starting from its most symmetric limit of a spin-$\tfrac{1}{2}$ Kondo impurity coupled isotropically to spin$-1$ conduction electrons, gradually breaking its symmetries, and then including the small nonlocal couplings not considered in Ref.~\onlinecite{Beri2012}. Using the NRG, we have obtained accurate results for the model, including directly-measurable quantities such as the differential conductance, on energy scales ranging over many orders of magnitude. 

The spin-isotropic limit of the model displays a universal crossover from local moment to non Fermi liquid fixed points, the latter being characterised by non-trivial power-law corrections to the low-temperature/frequency physics. Since the spin sector of the present model is the same as that of the four-channel Kondo model, many of the full crossover curves we have calculated here tend toward known asymptotes for the latter model.  On the other hand, quantities depending on the details of the charge and orbital sectors, such as the scattering t-matrix and differential conductance, show quite distinct physics. 

Focussing on the differential conductance, we showed that this has characteristic logarithmic tails at high energies, which cross over to power-law behavior with exponent $2/3$ at low energies,\cite{Beri2012} as shown in Fig.~\ref{fig:cond}. The NRG results show that the power law scaling sets in for energy scales $|\omega|\ll 10^{-2}\ T_K$ and the logarithmic tails require $|\omega|\gg 10^{3}\ T_K$. Depending on the value of the Kondo temperature, observing either or both of these asymptotes might be difficult in experiment. The universal scaling curve (in terms of $|\omega|/T_K$) obtained here by NRG, with its far broader domain than those of the asymptotes, is therefore indispensable for full quantitative comparisons to experiments. 

We find, as predicted in Ref.~\onlinecite{Beri2012}, that the physics of the model is robust to exchange anisotropy (which would arise physically due to differences between the tunnel couplings in the device). Furthermore, we found that the small and large $J_z$ regimes are related by a duality when $J_\perp$ is sufficiently weak. This duality is reminiscent of that seen in the two-channel Kondo model,\cite{Kolf2007} and enables analytical weak-coupling perturbative scaling results to be carried over to the large-$J_z$ regime.

Including non-local coupling between the Majoranas breaks the effective time-reversal symmetry of the model, causing a crossover ultimately to the strong coupling, Fermi liquid fixed point. As long as the perturbations are sufficiently weak that the energy scale $T^*_\text{FL}$ of the Fermi liquid crossover is $T^*_\text{FL}\ll T_K$, data collapse onto the scaling curve of the unperturbed problem for a wide range of energies $|\omega|> T_K$, see Fig.~\ref{fig:dyn_anotb}. In this case, the NFL to Fermi liquid crossover itself also has its own universal scaling curve in terms of $|\omega|/T^*_\text{FL}$, providing another signature of the NFL physics. 

For larger perturbations, of order $T_K$ and above, the NFL fixed point will not be approached so closely. But even in this situation, part of the slow crossover from the local moment to NFL fixed point could potentially be observed   in experiment, thus providing a signature of the topological Kondo effect---and hence the existence of Majorana fermions---in a real device.


\acknowledgements
This research was supported by EPSRC grants EP/I032487/1 (AKM,DEL) and EP/J017639/1 (NRC), the MC IEF and the Royal Society (BB).


\appendix

\section{Correlation functions for scattering t-matrix}
\label{sec:appt}
For a system described by the general Hamiltonian
\begin{equation}
\label{eq:genh}
\hat H = \sum_{k,m}\epsilon_k\ccre{k m}\cdes{k m} + \hat H_\mathrm{int},
\end{equation}
in which $\hat H_\mathrm{int}$ describes an interacting subsystem coupled to the $\cdes{k m}$ orbitals, and $m$ is a `flavor' index, standard equations of motion techniques lead to the result
\begin{multline}
\label{eq:tstart}
\langle\langle \cdes{k m};\ccre{k' m}\rangle\rangle = \frac{\delta_{kk'}}{z-\epsilon_k} + 
\frac{1}{(z-\epsilon_k)(z-\epsilon_{k'})} \times \\\left[\langle\{ [\cdes{k m}, \hat H_\text{int}], \ccre{k' m} \}\rangle +
\langle\langle [\cdes{k m}, \hat H_\text{int}]; [\cdes{k' m}, \hat H_\text{int}]^\dagger \}\rangle\rangle_\omega\right].  
\end{multline}
where $z = \omega + \I0^+$ and $\langle\langle \hat A;\hat B\rangle\rangle_\omega$ denotes the Fourier transform to the $\omega$ domain of the retarded correlation function $-\I\theta(t)\langle\{\hat A(t), \hat B\}\rangle$. The term in square brackets in \eref{eq:tstart} describes describes scattering due to $\hat H_\mathrm{int}$ and will be related to the on-shell t matrix below.

The Hamiltonian \eref{eq:heff} is of precisely the form of \eref{eq:tstart}, with $H_\mathrm{int}$ given by the terms involving the `impurity' spin operators. From \eref{eq:f-orb} we obtain $\{\cdes{k m},\fcre{m'}\} = N_\mathrm{orb}^{-1/2}\delta_{m m'}$, and use the relation $[\hat A,\hat B\hat C] = \{\hat A,\hat B\}\hat C - \hat B\{\hat A,\hat C\}$ to obtain
\begin{equation}
[\cdes{k m},\hat H_\mathrm{int}] = N_\mathrm{orb}^{-1/2} \hat X_m
\end{equation}
with
\begin{subequations}
\label{eq:xs}
\begin{align}
\hat X_1 &= \frac{J_\perp^a}{\sqrt{2}} S^+ \fdes{3} - J_z S^z \fdes{1}\\
\hat X_2 &= \frac{J_\perp^b}{\sqrt{2}} S^- \fdes{3} + J_z S^z \fdes{2}\\
\hat X_3 &= \frac{J_\perp^b}{\sqrt{2}} S^+ \fdes{2} + \frac{J_\perp^a}{\sqrt{2}} S^- \fdes{1}.
\end{align}
\end{subequations}
Defining $\tau_{k k',m}(\omega)$ as the quantity in square brackets in \eref{eq:tstart}, it follows from \eref{eq:xs} that the on-shell t-matrix is given by
\begin{align}
\tau_m(\omega) &= \sum_{k} \tau_{kk,m}(\omega)\\
 &= -J_z(\delta_{m 1} - \delta_{m 2}) \langle S_z \rangle + \langle\langle X_m^{\phantom{\dagger}}; X_m^\dagger \}\rangle\rangle_\omega
\end{align}
Of primary interest is the spectral function of $\tau_m(\omega)$, which we define as the dimensionless quantity
\begin{equation}
t_m(\omega) = - \pi \rho\, \mathrm{Im} \tau_m(\omega).
\end{equation}


\section{Kubo formula for conductance}
\label{sec:appk}
The conductance tensor of the model can be calculated for both the original basis of \eref{eq:model} and the transformed basis of \eref{eq:heff}. For the most part, the derivations are essentially identical, since only number operators of the lead electrons enter (rather than the interaction part of the Hamiltonian which is strongly basis-dependent). We therefore begin by outlining the part of the derivation common to both bases. 

Our starting point is the standard result from Appendix B of Hewson \cite{Hewson1997}. To leading order in the perturbation, the change in expectation value of an operator $\hat A$ from its equilibrium value, due to an adiabatically-switched-on perturbation $\hat H'(t) = \E^{\eta t}\hat B F(t)$ (with $\eta \to 0^+)$, is
\begin{equation}
\Delta \langle A(t)\rangle = \frac{1}{i\hbar}\mathrm{Tr} \int_{-\infty}^t \E^{\eta t'}[\hat B, \rho_\mathrm{eq}] \hat A(t-t') F(t')\D t' ,
\end{equation}
where $\Delta \langle A(t) \rangle = \mathrm{Tr}[\rho(t) \hat A - \rho_\mathrm{eq} \hat A]$, the density matrices in the presence [absence] of the perturbation are $\rho(t)$ [$\rho_\mathrm{eq}$], and 
\begin{equation}
\hat A(t-t') = \E^{\I\hat H (t-t')/\hbar} \hat A \E^{-\I \hat H(t-t')/\hbar}
\end{equation}
with $\hat H$ the Hamiltonian of the equilibrium system. Taking the perturbation to be
\begin{equation}
\label{eq:vdef}
\hat H'(t) = e \cos(\omega t) \hat N_\beta V_\beta
\end{equation}
(with $e$ the magnitude of the electron charge and $\omega$ the frequency of the AC bias voltage), the current into lead $\alpha$, 
\begin{equation}
\label{eq:idef}
I_\alpha(t) = e \langle \tfrac{\D}{\D t}{\hat N_\alpha}(t)\rangle \equiv e \langle \dot{N_\alpha}(t)\rangle
\end{equation}
is given by 
\begin{multline}
I_\alpha(t;\omega) =\frac{e^2 V_\beta}{\I\hbar}\\\times\mathrm{Tr} \int_{-\infty}^t \E^{\eta t'}[\hat N_\beta, \rho_\mathrm{eq}] \dot N_\alpha(t-t') \cos(\omega t')\D t'
\end{multline}
Here, we use $\hat N_\alpha$ to denote the total number operator of lead $\alpha$, either in the original basis of \eref{eq:model}, or the transformed basis of \eref{eq:heff}.

The change of variable $t-t' \to t''$ leads to 
\begin{multline}
I_\alpha(t;\omega) = \frac{e^2 V_\beta}{\I\hbar}\\\times\mathrm{Tr} \int_0^{\infty} \E^{-\eta t''}[\hat N_\beta, \rho_\mathrm{eq}] \dot N_\alpha(t'') \cos(\omega t'' - \omega t)\D t'',
\end{multline}
i.e.
\begin{equation}
I_\alpha(t;\omega) = -\frac{e^2 V_\beta}{2}\left[ \E^{-\I \omega t}\sigma_{\alpha\beta}(\omega) + \E^{\I \omega t}\sigma_{\alpha\beta}(-\omega)\right]
\end{equation}
with
\begin{equation}
\sigma_{\alpha\beta}(\omega) = \frac{i}{\hbar}\mathrm{Tr} \int_0^{\infty} \E^{\I(\omega+\I\eta) t}[\hat N_\beta, \rho_\mathrm{eq}] \dot N_\alpha(t)\D t.
\end{equation}
as defined by Izumida et al.\cite{Izumida1997} Following the manipulations therein, we obtain (for $\omega\ne 0$)
\begin{align}
\left(\frac{e^2}{h}\right)G_{\alpha\beta}(t;\omega) &= \frac{\partial I_\alpha(t;\omega)}{\partial V_\beta}\label{eq:gdefapp}\\
 &= -\left(\frac{e^2}{h}\right)2\pi\Bigl[\cos(\omega t) \frac{\hbar^2 \mathrm{Im} K^{\prime\prime}_{\alpha\beta}(\omega)}{\hbar\omega}
\nonumber \\&-\sin(\omega t) \frac{\hbar^2(K^{\prime}_{\alpha\beta}(\omega) - K^{\prime}_{\alpha\beta}(0))}{\hbar\omega}
\Bigr]
\end{align}
where 
\begin{align}
\label{eq:kdef}
K_{\alpha\beta}(\omega) &= -\frac{\I}{\hbar}\int_0^{\infty} \E^{\I(\omega+\I\eta) t}\left\langle\left[\dot N_\beta, \dot N_\alpha(t)\right]\right\rangle \D t \\
&= K^\prime_{\alpha\beta}(\omega) + \I K^{\prime\prime}_{\alpha\beta}(\omega).
\end{align}
In the $\omega\to 0$ limit, we obtain the steady-state ($t$-independent) DC conductance
\begin{equation}
\label{eq:gdcdef}
G^\mathrm{DC}_{\alpha\beta} \equiv G_{\alpha\beta}(t;\omega = 0) = 
 -2\pi \lim_{\omega \to 0}\frac{\hbar^2 K^{\prime\prime}_{\alpha\beta}(\omega)}{\hbar\omega}
\end{equation}
At finite $\omega$, for specificity we focus henceforth on the $t=0$ value of the conductance, 
\begin{equation}
\label{eq:gacdef}
G_{\alpha\beta}(\omega) \equiv G_{\alpha\beta}(t=0;\omega) = -2\pi \frac{\hbar^2 K^{\prime\prime}_{\alpha\beta}(\omega)}{\hbar\omega}
\end{equation}
which thus probes the frequency dependence of the same spectral function $K^\mathrm{\prime\prime}_{\alpha\beta}(\omega)$, and satisfies
\begin{equation}
\lim_{\omega\to 0} G_{\alpha\beta}(\omega) = G^\mathrm{DC}_{\alpha\beta}.
\end{equation}
By Lehmann-resolving $K_{\alpha\beta}(\omega)$ (see ref.~\onlinecite{Izumida1997}), one obtains
\begin{equation}
\mathrm{sgn}\left[K_{\alpha\alpha}^{\prime\prime}(\omega)\right] = \mathrm{sgn}(\omega)
\end{equation}
and hence, from \eref{eq:gacdef}, 
\begin{equation}
\label{eq:gsign}
G_{\alpha\alpha}(\omega) \le 0.
\end{equation}
To check this sign, one can examine the DC limit. Using \eref{eq:vdef} with $\omega\to 0$, a positive bias applied to lead $\alpha$ will raise its chemical potential, such that the current into lead $\alpha$ obtained from \eref{eq:idef} [and hence the conductance from \eref{eq:gdefapp}] should indeed be negative.


\subsection{Symmetries of the conductance}
For specificity we focus on a particular off-diagonal element, $G_{13}(\omega)$ of the conductance tensor. As explained below, in the axial-symmetric limit of the model it has the convenient properties of a) being the same in both bases, and b) being proportional also to the diagonal element $G_{33}(\omega)$.

We denote by $\hat{\mathcal{N}}_\alpha$ and $\hat N_m$ the number operators for the leads in the physical basis of \eref{eq:model} and the transformed basis of \eref{eq:heff}, respectively. It is straightforward to show that  
\begin{subequations}
\begin{align}
\hat N_{-1} + \hat N_1 &= \hat{\mathcal N}_1 + \hat{\mathcal N}_2\\
\hat N_0 &= \hat{\mathcal N}_3.
\end{align}
\end{subequations}
It thus follows from eqns.~(\ref{eq:kdef}) and (\ref{eq:gacdef}) that 
\begin{subequations}
\label{eq:gsym}
\begin{align}
G_{10}(\omega) + G_{-10}(\omega) &= \mathcal{G}_{13}(\omega) + \mathcal{G}_{23}(\omega)\\
G_{00}(\omega) &= \mathcal{G}_{33}(\omega).
\end{align}
\end{subequations}
where $\mathcal{G}_{\alpha\beta}(\omega)$ and $G_{nm}(\omega)$ denote the conductance in  the physical and transformed bases, respectively.

Two sets of further results hold in each basis: axial symmetry implies that 
\begin{subequations}
\label{eq:gax}
\begin{align}
G_{10}(\omega) &= G_{-10}(\omega),\\
\mathcal{G}_{13}(\omega) &= \mathcal{G}_{23}(\omega),
\end{align}
\end{subequations}
while current conservation means that
\begin{subequations}
\label{eq:gcc}
\begin{align}
G_{-10}(\omega)+G_{00}(\omega)+G_{10}(\omega) &= 0,\\
\mathcal{G}_{13}(\omega)+\mathcal{G}_{23}(\omega)+\mathcal{G}_{33}(\omega) &= 0.
\end{align}
\end{subequations}
Combining eqns.~(\ref{eq:gsym}), (\ref{eq:gax}) and (\ref{eq:gcc}), we obtain
\begin{equation}
\label{eq:gsymms}
G_{10}(\omega)= -\tfrac{1}{2} G_{00}(\omega) 
 = \mathcal{G}_{13}(\omega)= -\tfrac{1}{2}\mathcal{G}_{33}(\omega) \ge 0.
\end{equation}
[Here, the minus signs reflect that, for a particular bias applied to lead $0$, the currents flowing from the `impurity' to leads $1$ and $0$ will be in opposite directions, and the inequality is obtained from \eref{eq:gsign}.] The single quantity $G_{10}(\omega)$ thus provides a useful handle on the conductance of the model in both bases.

In the fully isotropic limit of the model, \eref{eq:gsymms} can of course be extended by symmetry: the full tensors $G$ and $\mathcal{G}$ are identical, with 
\begin{equation}
G_{mn}(\omega)= -\tfrac{1}{2}G_{mm}(\omega) = -\tfrac{1}{2} G_{nn}(\omega)
\end{equation}
for all $m\ne n$. In this case, therefore, the single quantity $G_{13}(\omega)$ represents all elements of both conductance tensors.


\subsection{Calculation via NRG}
To calculate $G_{10}(\omega)$ by NRG, one needs the time-derivatives of the lead operators in \eref{eq:kdef}. Using 
\begin{align}
\dot N_m &= \frac{1}{i\hbar}\left[\hat N_m, \hat H\right],
\end{align}
one obtains $\dot N_m = \hat o_m / (i\hbar)$ with
\begin{align}
\hat o_{-1} &= \frac{J_\perp^a}{\sqrt{2}}\left(S^+ \fpair{-1}{0} - S^- \fpair{0}{-1}\right) \\
\hat o_{1} &= \frac{J_\perp^b}{\sqrt{2}}\left(S^- \fpair{1}{0} - S^+ \fpair{0}{1}\right) \\
\hat o_0 &= - \hat o_{-1} - \hat o_1 
\end{align}
(the latter equality reflecting the current conservation in the system). 

Matrix elements of these operators can be calculated, and then used to obtain the correlation function $K_{mn}$ in \eref{eq:kdef} by the FDM-NRG method.\cite{Weichselbaum2007} Owing to the discretization inherent in the NRG approach, one obtains a discrete set of delta functions which lead to a similarly discrete representation of $G_{mn}(\omega)$ when \eref{eq:gacdef} is used. We then broaden the poles of $G_{mn}(\omega)$ using the standard method of Ref.~\onlinecite{Weichselbaum2007}.


%

\end{document}